\documentclass[twocolumn,pre,floatfix]{revtex4}
\usepackage{graphicx}
\usepackage{tabularx}
\usepackage{epsfig}
\usepackage{rotating}
\usepackage{amsmath}
\usepackage{amsfonts}
\usepackage{amssymb}
\usepackage{enumerate}
\usepackage{longtable}
\usepackage{empheq,color,cancel,setspace}
\usepackage{dcolumn}
\usepackage{bm}
\usepackage{tikz}
\newcommand{\chiSG}{\chi_{\raisebox{-0.6ex}{$\scriptstyle SG$}}}

\begin{document}

\title{Efficient generation of series expansions for $\pm J$
Ising spin-glasses in a classical or a quantum (transverse) field}

\author{R. R. P. Singh}
\affiliation{Physics Department, University of California Davis, CA 95616, USA}

\author{A. P. Young}
\affiliation{Physics Department, University of California Santa Cruz, CA 95064, USA}

\date{\rm\today}

\begin{abstract}
We discuss generation of series expansions for Ising spin-glasses with a
symmetric $\pm J$ (i.e.~bimodal) distribution on d-dimensional hypercubic
lattices using linked-cluster methods.  Simplifications for
the bimodal distribution allow us to go to higher order than for a
general distribution. We discuss two types of problem, one classical and one
quantum.  The classical problem is that of the Ising spin glass in a
longitudinal magnetic field, $h$, for which we obtain high temperature series
expansions in variables $\tanh(J/T)$ and $\tanh(h/T)$.  The quantum problem is
a $T=0$ study of the Ising spin glass in a transverse magnetic field $h_T$ for
which we obtain a perturbation theory in powers of $J/h_T$. These methods
require (i) enumeration and counting of \textit{all} connected clusters that can be
embedded in the lattice up to some order $n$, and (ii) an evaluation of the
contribution of each cluster for the quantity being calculated,
known as the weight.
We discuss a general method that takes the much
smaller list (and count) of all no free-end (NFE) clusters on a lattice up to
some order $n$, and automatically generates all other clusters and their counts
up to the same order. The weights for finite clusters in both
cases have a simple graphical interpretation that allows us to proceed
efficiently for a general configuration of the $\pm J$ bonds, and at the end
perform suitable disorder averaging. The order of our computations is limited
by the weight calculations for the high-temperature expansions of the
classical model, while they are limited by graph counting for the $T=0$
quantum system. Details of the calculational methods are presented.

\end{abstract}

\maketitle

\section{Introduction}
\label{sec:intro}
The controlled study of short-range Ising spin-glass models on finite dimensional
lattices is well known to be a challenging
task~\cite{edwards:75,sherrington:75,binder:86}. Important problems include the
question of the de Almeida Thouless instability in a magnetic field
~\cite{almeida:78}, and the quantum critical behavior and Griffiths-McCoy
singularities in the presence of a transverse quantum field at 
$T=0$~\cite{fisher:92,fisher:95,griffiths:69,mccoy:69,mccoy:69b,vojta:10}.
For both these problems it is of great interest to investigate the behavior as
a function of dimension.

One of the most successful methods for studying spin glasses is Monte
Carlo simulations \cite{binder:86}.  Advances in simulation methods have led
to substantial insights especially when the dimensionality is not too high. Here
we consider an alternative approach, series expansions. Whereas it is
difficult to study spin glasses in high dimensions, because the range of
(linear) sizes which can be studied is too limited to perform a satisfactory
finite-size scaling, the complexity of the series method depends only weakly
on dimensionality\cite{fisch:77,singh:86,fisher:90,klein:91}, so it is particularly useful for the
study of spin glasses in high dimensions. Another advantage of series
expansions is that the average over disorder is done exactly. 
The purpose of this work
is to describe an efficient method for calculating such series expansions.
The results of our calculations for classical~\cite{singh:17b} and quantum
systems~\cite{singh:17} have been published elsewhere.
 
\section{The Linked Cluster Method for Series Expansions}
\label{sec:linked_cluster}

\subsection{The basic idea}
Our common framework for developing series expansions for both classical and
quantum systems is the Linked Cluster method \cite{oitmaa:06,gelfand:90}. In this method, we consider an
extensive property of interest $P$ and compute
$\lim_{N\to\infty} P/N$, where $N$ is the
number of sites of the lattice, i.e.~ the property $P$ \textit{per site} in the
thermodynamic limit. We expand this quantity in powers of a suitable expansion
variable $x$.
The essence of the Linked Cluster methods is to express $P/N$ as \cite{oitmaa:06,gelfand:90}
\begin{equation}
P/N = \sum_c L(c) \times W(c),
\end{equation}
where the sum is over all distinct \textit{connected} clusters that can be embedded in the
lattice. The quantity $L(c)$ is called the lattice constant of the cluster
$c$. It is the number of ways the cluster $c$ can be embedded in the lattice
\textit{per lattice site}. The quantity $W(c)$ is called the weight of the cluster $c$
and is given by the recursive relation
\begin{equation}
W(c) = P(c) -\sum_{s\subset c} W(s),
\end{equation}
where $P(c)$ is the value of property $P$ evaluated for cluster
$c$~\cite{intensive}, and the
sum over $s$ is over all ``proper'' 
sub-clusters of the cluster $c$ (i.e.~the sum excludes $c$ itself).
Thus a series calculation requires:
\begin{enumerate}
\item
enumeration and counting of all relevant
clusters, and 
\item
calculation of the weight of each cluster, which needs to be
expanded as a power series in the expansion variable $x$.
\end{enumerate}
The weight of a
cluster with $n$ bonds can be shown to be of
order \cite{oitmaa:06,gelfand:90} $x^n$. Thus, summing up
contributions from all clusters with $n$ or fewer bonds gives the
series expansion for $P/N$ to order $n$.

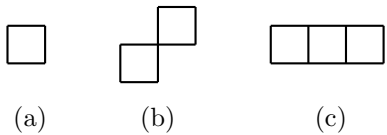
\begin{figure}
\begin{tikzpicture}
\node  (a) at (0.3,-0.75) { (a)};
\draw [thick] (0,0) -- (0.5,0);
\draw [thick]  (0.5, 0) -- (0.5,0.5);
\draw [thick]  (0.5,0.5) -- (0,0.5);
\draw [thick]  (0,0.5) -- (0,0);

\node  (b) at (2.0,-0.75) {(b)};
\draw [thick] (1.5,-0.25) -- (2.0,-0.25);
\draw [thick]  (2.0, -0.25) -- (2.0,0.25);
\draw [thick]  (2.0,0.25) -- (1.5,0.25);
\draw [thick]  (1.5,0.25) -- (1.5,-0.25);
\draw [thick] (2.0,0.25) -- (2.5,0.25);
\draw [thick]  (2.5, 0.25) -- (2.5,0.75);
\draw [thick]  (2.5,0.75) -- (2.0,0.75);
\draw [thick]  (2.0,0.75) -- (2.0,0.25);

\node  (c) at (4.3,-0.75) {(c)};
\draw [thick]  (3.5,0) -- (3.5,0.5);
\draw [thick] (4.5,0.5) -- (5,0.5);
\draw [thick]  (5.0, 0.5) -- (5.0,0);
\draw [thick]  (3.5, 0) -- (5.0,0);
\draw [thick]  (4.5, 0) -- (4.5,0.5);
\draw [thick]  (4.5, 0.5) -- (3.5,0.5);
\draw [thick]  (4.0, 0.5) -- (4.0,0);
\end{tikzpicture}
\caption{\label{fig:NFE}
Three examples of graphs with no free ends. Of these, (a) and (c) are
also star graphs, but (b) is not because it can be cut in two by removing the
vertex where the two squares join.
}
\end{figure}

For certain problems such as classical spin-glasses in zero-field, many
quantities only require a limited type of clusters. One of the most efficient
such method is a star-graph expansion~\cite{ditzian:79,singh:86} that requires only clusters that do not
have articulation points. An articulation point is a node which, if removed,
would split the graph into disconnected pieces.
Certain other calculations \cite{oitmaa:06,fisch:77,klein:91}
may only require clusters with
zero or at most two free ends (a free end is a site with only one other site
connected to it). Figure \ref{fig:NFE} shows several clusters with no free
ends (NFE). Clearly all star graphs, except the trivial one with one bond,
are of the NFE type, whereas some NFE graphs are not stars, an example being
cluster (b) in Fig.~\ref{fig:NFE}.
However, the problems we study here with classical and quantum fields require
consideration of \textit{all} connected clusters. Since this part of the calculation
is common to both classical and quantum study we discuss this enumeration and
counting problem first. We then separately consider the weight calculations
for the quantum and classical systems.

\subsection{Graph enumeration and lattice constants}
\label{sec:graphs}
We consider nearest-neighbor models on d-dimensional hypercubic lattices.
Furthermore, we will only consider properties such as spin-glass
susceptibility where the weights of a cluster only depend on the connectivity
(or adjacency matrix) of the cluster and not on the many ways in which the
cluster may be embedded in the $d$-dimensional lattice. Hence we will not be
able to calculate, for example, the correlation length which requires a
knowledge of the vector between each pair of sites in the cluster. In this
section we will use the terms cluster and graph interchangeably to mean the
same thing.

Following Fisher and Gaunt \cite{fisher:64}, the
lattice constant for such a cluster $c$ in dimension $d$ can be expressed as
\begin{equation}
L_d(c) = \sum_{m=m_{min}}^{m_{max}} {d \choose m} l_m(c),
\label{gen_d}
\end{equation}
where ${d \choose m}$ is the binomial coefficient.
Here $l_m(c)$ is the count of those embeddings of the cluster $c$ that extend
in $m$ dimensions. The limits of the summation $m_{min}$ and $m_{max}$
refer to the minimum and maximum dimension in which the cluster embeddings can
extend. Clearly a cluster with $n$ bonds can  not extend in more than $d=n$
dimensions, but if there are closed loops then $m_{max}$ is less than $n$.
Using Eq.~\eqref{gen_d}, one can
go back and forth between $L_d(c)$ and $l_m(c)$. One first calculates $L_d(c)$
in different dimensions up to the maximum dimension where there is an
embedding of cluster $c$ and then uses Eq.~\eqref{gen_d} to obtain $l_m(c)$.
Once that is done, one can now readily obtain the lattice constants in
arbitrary dimension by using Eq.~\eqref{gen_d} again. For the rest of this
section we will assume that we are working with a fixed dimensionality $d$.

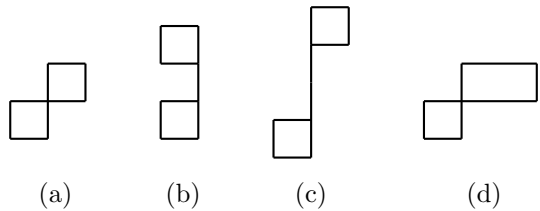
\begin{figure}
\begin{tikzpicture}

\node  (a) at (-1.4,-0.75) { (a)};
\draw [thick] (-2,0) -- (-1.5,0);
\draw [thick]  (-1.5, 0) -- (-1.5,0.5);
\draw [thick]  (-1.5,0.5) -- (-2,0.5);
\draw [thick]  (-2,0.5) -- (-2,0);
\draw [thick] (-1.5,0.5) -- (-1.5,1.0);
\draw [thick]  (-1.5, 1.0) -- (-1.0,1.0);
\draw [thick]  (-1.0,1.0) -- (-1.0,0.5);
\draw [thick]  (-1.0,0.5) -- (-1.5,0.5);

\node  (b) at (0.3,-0.75) { (b)};
\draw [thick] (0,0) -- (0.5,0);
\draw [thick]  (0.5, 0) -- (0.5,0.5);
\draw [thick]  (0.5,0.5) -- (0,0.5);
\draw [thick]  (0,0.5) -- (0,0);
\draw [thick]  (0.5,0.5) -- (0.5,1.0);
\draw [thick] (0,1.0) -- (0.5,1.0);
\draw [thick]  (0.5, 1.0) -- (0.5,1.5);
\draw [thick]  (0.5,1.5) -- (0,1.5);
\draw [thick]  (0,1.5) -- (0,1.0);

\node  (c) at (2.0,-0.75) {(c)};
\draw [thick] (1.5,-0.25) -- (2.0,-0.25);
\draw [thick]  (2.0, -0.25) -- (2.0,0.25);
\draw [thick]  (2.0,0.25) -- (1.5,0.25);
\draw [thick]  (1.5,0.25) -- (1.5,-0.25);
\draw [thick]  (2.0,0.25) -- (2.0,0.75);
\draw [thick]  (2.0,0.75) -- (2.0,1.25);
\draw [thick] (2.0,1.25) -- (2.5,1.25);
\draw [thick]  (2.5, 1.25) -- (2.5,1.75);
\draw [thick]  (2.5,1.75) -- (2.0,1.75);
\draw [thick]  (2.0,1.75) -- (2.0,1.25);

\node  (d) at (4.3,-0.75) {(d)};
\draw [thick]  (3.5,0) -- (3.5,0.5);
\draw [thick]  (3.5,0) -- (4.0,0);
\draw [thick]  (4.0,0) -- (4.0,0.5);
\draw [thick]  (3.5,0.5) -- (5,0.5);
\draw [thick]  (5.0, 0.5) -- (5.0,1.0);
\draw [thick]  (5.0, 1.0) -- (4.0,1.0);
\draw [thick]  (4.0, 1.0) -- (4.0,0.5);
\end{tikzpicture}
\caption{\label{fig:NFE10}
A complete list of non-star NFE graphs (a) through (d) that can be embedded in
hypercubic lattices with $10$ or fewer bonds.
}
\end{figure}

\begin{table}[h]
\caption{\label{Tab:non-star}Counts of embeddings $l_m(c)$ of the non-star, NFE clusters in
Fig. \ref{fig:NFE10} that extend in exactly $m$ dimensions. Note that none of
these NFE clusters extend beyond $6$ dimensions.
}
\begin{tabular}{|c|r|r|r|r|r|}
\hline
\ cluster in Fig. \ref{fig:NFE10} \   & $m=2$ & $m=3$ & $m=4$ & $m=5$ & $m=6$ \\
\hline\hline
(a)  & 2 & 24 & 48 & 0 & 0 \\
(b)  & 8 & 168 & 576 & 480 & 0 \\
(c)  & 20 & 792 & 5184 & 10080 & 5760 \\
(d)  & 8 & 552 & 3168 & 3840 & 0 \\
\hline
\end{tabular}
\end{table}

We will now describe a method to obtain all connected clusters up to order 10
in any dimension starting from the NFE clusters and their lattice constants.
The lattice constants of star-graphs to order $11$ in
general dimension were provided by Ditzian and Kadanoff \cite{ditzian:79}.
According to their Table III,
there are only $17$ star-graphs with $10$ or less bonds.
In addition, there
are $4$ non-star, NFE graphs up to order 10, and these are shown in
Figure \ref{fig:NFE10} while their lattice constants are given in
Table~\ref{Tab:non-star}. Hence the number of NFE graphs up to order 10 is 21.
By
contrast, we find that the \textit{total} number of graphs with $10$ or less bonds
increases to $933$ when free-ends are allowed. Furthermore, general graphs
with $10$-bonds can be embedded in all dimensions up to $d=10$, and their
lattice constants can become enormous. To get an idea, up to $10$ bonds the
largest lattice constant for an NFE graph for $d\le 10$
is about $10^7$.
By contrast, lattice constant for graphs with free ends
can be of order $10^{13}$ for $d\le 10$. Thus a \textit{direct} enumeration
of all embeddings of general graphs becomes very challenging
especially in high dimensions.

Instead, we will develop here a method which generates a list of all graphs
(including those with free ends) and their lattice constants from the lattice
constants of graphs with no free-ends, without any further \textit{explicit}
enumeration of the embeddings.

It is evident from the meaning of a free-end that any graph with $n+1$ bonds
that has at least one free-end can be obtained from some graph with $n$ bonds
by attaching a bond to one of its sites. Such an $n$-bond graph can be
obtained by just cutting off the last bond on one of the free-ends.

\begin{figure}
\begin{tikzpicture}
\filldraw[black] (-4.0,2) circle (1pt);
\filldraw[black] (-3.5,2) circle (1pt);
\filldraw[black] (-3.0,2) circle (1pt);
\filldraw[black] (-2.5,2) circle (1pt);
\filldraw[black] (-2.0,2) circle (1pt);
\filldraw[black] (-1.5,2) circle (1pt);
\filldraw[black] (-1.0,2) circle (1pt);
\filldraw[black] (-0.5,2) circle (1pt);
\draw [thick] (-4,2.0) -- (-0.5,2.0);
\node  (a) at (-2.25,1.25) {(a)};

\filldraw[black] (4.0,2) circle (1pt);
\filldraw[black] (3.5,2) circle (1pt);
\filldraw[black] (3.0,2) circle (1pt);
\filldraw[black] (2.5,2) circle (1pt);
\filldraw[black] (2.0,2) circle (1pt);
\filldraw[black] (1.5,2) circle (1pt);
\filldraw[black] (1.0,2) circle (1pt);
\filldraw[black] (0.5,2) circle (1pt);
\filldraw[black] (0.0,2) circle (1pt);
\draw [thick] (0.0,2.0) -- (4.0, 2.0);
\node  (b) at (2.0,1.25) {(b)};

\filldraw[black] (-4.5,0.25) circle (1pt);
\filldraw[black] (-4.0,0.25) circle (1pt);
\filldraw[black] (-3.5,0.25) circle (1pt);
\filldraw[black] (-3.0,0.25) circle (1pt);
\filldraw[black] (-4.5,-0.25) circle (1pt);
\filldraw[black] (-4.0,-0.25) circle (1pt);
\filldraw[black] (-3.5,-0.25) circle (1pt);
\filldraw[black] (-3.0,-0.25) circle (1pt);
\draw [thick] (-4.5,0.25) -- (-3.0, 0.25);
\draw [thick] (-3.0,0.25) -- (-3.0, -0.25);
\draw [thick] (-3.0,-0.25) -- (-4.5, -0.25);
\draw [thick] (-4.5,-0.25) -- (-4.5, 0.25);
\node  (c) at (-3.75,-1) {(c)};

\filldraw[black] (-1.5,0.25) circle (1pt);
\filldraw[black] (-1.0,0.25) circle (1pt);
\filldraw[black] (-0.5,0.25) circle (1pt);
\filldraw[black] ( 0.0,0.25) circle (1pt);
\filldraw[black] ( 0.5,0.25) circle (1pt);
\filldraw[black] ( 1.0,0.25) circle (1pt);
\filldraw[black] (-1.5,-0.25) circle (1pt);
\filldraw[black] (-1.0,-0.25) circle (1pt);
\draw [thick] (-1.5,-0.25) -- (-1.0,-0.25);
\draw [thick]  (-1.0, -0.25) -- (-1.0,0.25);
\draw [thick]  (-1.0,0.25) -- (-1.5,0.25);
\draw [thick]  (-1.5,0.25) -- (-1.5,-0.25);
\draw [thick] (-1.0,0.25) -- (1.0,0.25);
\node  (d) at (-0.25,-1) {(d)};

\filldraw[black] (2.0,0.25) circle (1pt);
\filldraw[black] (2.5,0.25) circle (1pt);
\filldraw[black] (3.0,0.25) circle (1pt);
\filldraw[black] (3.5,0.25) circle (1pt);
\filldraw[black] (4.0,0.25) circle (1pt);
\filldraw[black] (2.0,-0.25) circle (1pt);
\filldraw[black] (2.5,-0.25) circle (1pt);
\filldraw[black] (3.0,-0.25) circle (1pt);
\draw [thick] (2.0,-0.25) -- (3.0,-0.25);
\draw [thick]  (3.0, -0.25) -- (3.0,0.25);
\draw [thick]  (3.0,0.25) -- (2.0,0.25);
\draw [thick]  (2.0,0.25) -- (2.0,-0.25);
\draw [thick] (3.0,0.25) -- (4.0,0.25);
\node  (e) at (3.0,-1) {(e)};

\end{tikzpicture}
\caption{\label{fig:tadpole}
Various graphs arising in Eq.~\eqref{example0}:
(a) linear graph of length $7$ denoted $c_7$,
(b) linear graph of length $8$ denoted $c_8$,
(c) polygon of length $8$ denoted $p_8$,
(d) tadpole graph with one free-end denoted $t_{4,4}$,
(e) tadpole graph with one free-end
denoted $t_{6,2}$. 
}
\end{figure}
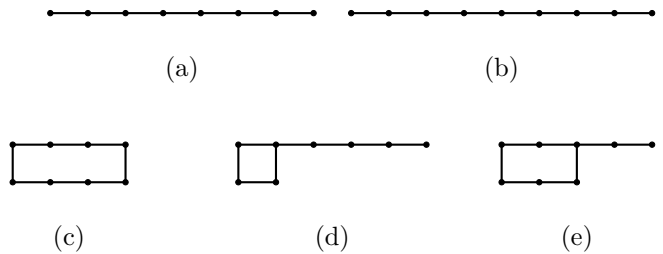

To generate the lattice constants of these free-end graphs,  we will use a
method along the lines of early work of Domb \cite{domb:60,domb:73} and Fisher
and Gaunt \cite{fisher:64}.  These authors showed that there are relationships
between lattice constants of different graphs. For example, if we take a chain
of length $n$ and attach to it an additional bond at the end, it will create a
chain of length $n+1$. Thus the lattice constant for a chain of length $n+1$
should be related to the lattice constant for a chain of length $n$ times the
number of ways a bond can be added to the ends, except that upon addition of
the bond, it can touch one of the existing sites of the chain thus forming
either a closed loop of length $n+1$ or a tadpole graph
\cite{domb:60,fisher:64}, see Fig.~\ref{fig:tadpole}.

For
concreteness consider $n=7$. Let us call the length $7$ chain as $c_7$, length
$8$ chain as $c_8$, the polygon of length $8$ as $p_8$, a tadpole with a length
$2$ chain attached to a loop of length $6$ as $t_{6,2}$, and a tadpole with a
length $4$ chain attached to a loop of length $4$ as $t_{4,4}$. These graphs
are shown in Fig.~\ref{fig:tadpole}. On a
d-dimensional lattice there are $2d-1$ ways to add a bond at either end of a
chain. Hence, the relationship is:
\begin{eqnarray}
\label{example0}
2 (2d-1) L_d(c_7) =& 2 L_d(c_8) + 2\times 8 L_d(p_8) \\ \nonumber 
                   &+ 2 L_d(t_{6,2}) + 2 L_d(t_{4,4}),
\end{eqnarray}
where the factors in front of the lattice constants are related to the
symmetries of the graphs. The factor of 2 on the left hand side is related to
the fact that the bond can be added on either end. The factor of 2 before
$L_d(c_8)$ is because the same embedding of $c_8$ can arise in two ways by
addition of last bond on left or right. The factor of $2\times 8$ for
$l_d(p_8)$ is because any one of the $8$ bonds of the polygon could have been
added to a chain to form the polygon and it could be added from either end.
The factor of $2$ for the other two graphs is because the last bond added that
forms the tadpole from a chain of length $7$ must be one of the two bonds
inside the loop next to the site with valency $3$. Such relations exist
whenever a bond is added to a graph.

Before we discuss our method further, we need to define a graph theoretic
property called the cyclomatic number.  The cyclomatic number of a graph is
defined by the number of independent cycles in the graph \cite{domb:73}. It
equals $n-m+1$, where $n$ is number of bonds and $m$ is number of sites. Thus
a tree graph, which has one less bond than sites, 
has a cyclomatic number of $0$ in agreement with there being no cycles.
The important thing to note about the above 
Eq.~\eqref{example0} is that only one graph on the right hand side, the first
one, has the same cyclomatic number as the graph on the left hand side. All
other graphs have one higher cyclomatic number as an additional loop has been
formed. Thus, if the lattice constants of all graphs with higher cyclomatic
number are already known, this equation can help us determine $L_d(c_8)$ from
$L_d(c_7)$.

We will use the facts that (i) all graphs with free-ends can be obtained by
adding a bond to some other graph with one less bond, (ii) lattice constants
of graphs can be related by some relation such as in Eq.~\eqref{example0}, and
(iii) highest cyclomatic number graphs in any order must either have no
free-ends or can be obtained by adding a bond to a smaller no free-end graph
with the same cyclomatic number, without the creation of any additional graphs
on the right hand side of Eq.~\eqref{example0}.  Using these facts, and working in
order of decreasing cyclomatic number, we next describe an algorithm that allows us
to obtain a list of all graphs and their lattice constants.

We work in order of increasing number of bonds, and,
for a given number of bonds, we work in order of decreasing cyclomatic number.
Suppose a complete list of all graphs in order $n$ is already available. Then,
we obtain the complete list of all graphs in order $n+1$ by adding bonds to
all $n$ bond graphs in all possible ways, plus the additional NFE graphs of
order $n+1$, whose counts already exist.  Note that in such a scheme the same
graph can be generated many times and the duplicates have to be removed using
standard methods \cite{oitmaa:06,gelfand:90}.  Then the only thing that
remains is find the analog of Eq.~\eqref{example0} every time a new graph is
found. We do this by an automated computer algorithm discussed next. This
automated algorithm for recognizing the desired equation from which lattice
constant of the new graph can be read off is one of the most important
developments in this paper.

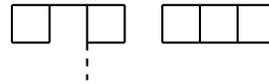
\begin{figure}
\begin{tikzpicture}
\draw [thick] (0,0) -- (0.5,0);
\draw [thick]  (0.5,0) -- (0.5,0.5);
\draw [thick] (0.5,0.5) -- (0,0.5);
\draw [thick]  (0, 0.5) -- (0,0);
\draw [thick]  (0.5, 0.5) -- (1.5,0.5);
\draw [thick]  (1.5, .5) -- (1.5,0);
\draw [thick]  (1.5, 0) -- (1,0);
\draw [thick]  (1, 0) -- (1,0.5);
\draw [thick,dashed] (1,0) -- (1,-0.5);

\draw [thick] (2,0) -- (2.5,0);
\draw [thick]  (2.5,0) -- (2.5,0.5);
\draw [thick] (2.5,0.5) -- (2,0.5);
\draw [thick]  (2.0, 0.5) -- (2.0,0);
\draw [thick]  (2.5, 0) -- (3.5,0);
\draw [thick]  (3.5, 0) -- (3.5,0.5);
\draw [thick]  (3.5, 0.5) -- (2.5,0.5);
\draw [thick]  (3.0, 0.5) -- (3.0,0);
\end{tikzpicture}
\caption{\label{fig:parent} 
Example of a parent graph, $g_p$ (on the left without the dashed line), a
daughter graph, $g_d$, on the left with the dashed line added, and additional
graph $g_1$ (on the right) needed to get a closed equation for the lattice
constants.
}
\end{figure}

Consider a parent graph $g_p$ with $n$ bonds and $m$ sites. We generate a
daughter graph $g_d$ with $n+1$ bonds and $m+1$ sites, either with an extra
free end or with the length of one of the free ends extended by one
(see Fig.~\ref{fig:parent}), by (a) picking a site $i$ of the parent graph and
(b) addding a bond connecting
this site to an additional site, which we label as $m+1$.

We would like to find an equation that relates the
lattice constant of the daughter graph to the lattice constant of the parent
graph. In order to
have such a relation, we need to know what other graphs can result in the
process of adding a bond from site $i$ to one of the existing sites
of the graph.  We can find a list of all such graphs by attempting to add a
bond between site $i$ and any other site of the graph and see if that forms an
allowed graph with one additional loop.
Let us remember that when graphs are generated in order
of decreasing cyclomatic number (decreasing number of loops) then counts for
such graphs will already be known.  Let us say we obtain
$q$ such graphs we can label 1 through $q$ and let the valency of the site $i$ in
the parent graph be $v$. Then the desired equation is
\begin{equation}
L_d(g_d)={ (2d-v)s_pL_d(g_p)-\sum_{n=1}^qs_nL_d(g_n)\over s_d}
\label{example-2}
\end{equation}
Here, $s_p$ is the number of equivalent sites in the parent graph where
addition of a bond also leads to same daughter graph, $s_d$ is the number of
ways the daughter graph can be generated from the parent graph and $s_n$ is
the number of ways the graph $g_n$ gets generated by adding a bond from a site
of the parent graph equivalent to $i$ to another site. All these factors $v$,
$s_d$, $s_p$ and $s_n$ depend only on the connectivity or adjacency matrix of
the graph and not its actual embeddings in a particular lattice. Hence, they
can be calculated by an automated computer algorithm from the adjacency matrix
of the graph.  Thus, the lattice constant of the graph $g_d$ can be obtained.

We have already seen one example of such a relation
in Fig.~\ref{fig:tadpole} and Eq.~\eqref{example0}.
Another example is shown in
Fig.~\ref{fig:parent}. The parent graph is a $9$-bond, $8$-site graph
with no free ends. Adding a bond as shown in the figure as a protruding dashed
line, produces a $10$-bond, $9$-site daughter graph with one free end.  In
this case $v=2$, $s_p=4$ and $s_d=1$. There is only one other type of graph that can
result by connecting the site under consideration to one of the existing sites
in the parent graph. That is a $10$-bond $8$-site graph also shown in
Fig.~\ref{fig:parent}.  In this problem $s_n=4$ as there are two bonds whose
removal can lead to the parent graph from the daughter graph and at either end
of the bond the site becomes equivalent to our addition site in the parent
graph. So, the equation becomes
\begin{equation}
L_d(g_d)= 4(2d-2)L_d(g_p)-\  4L_d(g_1)
\label{example-2a}
\end{equation}

When this procedure is followed starting with a single-bond graph and its
count, which is $d$, together with a list of all NFE graphs, a complete list
and count of all graphs results.

For the Sherrington-Kirkpatrick~\cite{sherrington:75} model, every spin  
interacts with every other spin and the variance of the interactions is
proportional to $1/N$ to get a sensible thermodyamic limit. In our work we
obtain this model by taking the limit $d \to \infty$ and scaling the variance
of the interactions by $1/(2 d)$ (the inverse of the number of neighbors).
It is well known that only tree graphs are needed in this limit.
Since NFE
graphs all have at least one closed loop, their counts become negligible together 
with those of all graphs with at least one loop,
relative to the tree graphs. The relevant count of these tree graphs is
simply given by the largest power of $d$ that arises in Eq. \eqref{gen_d}.
To obtain a
list and count of just the tree graphs, one can start with a single bond graph and carry out the procedure
discussed above ignoring any graphs with loops. In that case, equations 
such as Eq.~\eqref{example-2} have only
the parent and daughter graph and no additional graphs. One can show that, the factors of $s_d$ and $s_p$ in the
Eq.~\eqref{example-2} lead to a count of these tree graphs with $n$-bonds that
can be expressed as
\begin{equation}
L_{SK}(c) = {(2d)^n\over p_c},
\label{SK-count}
\end{equation}
where $p_c$ is the symmetry factor of the graph, defined by the number of permutations of the sites of the graph
that leaves the graph invariant. The factor of $(2 d)^n$ then cancels with
the $n$ factors of the variance of the bonds, leading to a result which is
independent of $d$ for $d \to \infty$. 

We have generated graphs to $14$th order for SK model, to $10$th order for
general $d$ and also to $14$th order in $d=2$ and $d=3$, where NFE graph
counts had been previously generated by an explicit enumeration
\cite{devakul}.  Thus, graph counting is currently limited by the availability
of counts of NFE graphs.

\section{Finite Cluster Calculations for quantum Ising spin glasses}
\subsection{Some Preliminaries}

For the transverse field Ising spin glass calculations, the unperturbed
Hamiltonian is the transverse field so we work in a representation in which
this is diagonal, rather than the usual representation in which the Ising spin
glass part is diagonal.
The Hamiltonian is therefore
\begin{equation}
\mathcal{H} = \mathcal{H}_0 + \mathcal{H}_1
\label{ham}
\end{equation}
where the unperturbed Hamiltonian is 
\begin{equation}
\mathcal{H}_0 = - \sum_k \sigma^z_k \, ,
\end{equation}
in which we have set the transverse field, $h^T$, equal to $1$, 
and the perturbation is the Ising spin glass part,
\begin{equation}
\mathcal{H}_1 = \sum_{\langle j, k\rangle} J_{jk} \sigma^x_j \sigma^x_k \, .
\end{equation}

In this section we
consider a single cluster with $N$ sites and $B$ bonds. 
The interactions take values 
\begin{equation}
J_b = \epsilon_b J\, ,\ \text{with}\ \epsilon_b = \pm 1, 
\label{Jb}
\end{equation}
where ($b = 1, 2, \cdots B$). Averages over disorder are simple with this
bimodal distribution since
\begin{equation}
J_b^n = \left\{
\begin{array}{ll}
J^n, & (n\ \text{even}), \\
\epsilon_b\, J^n, & (n\ \text{odd}),
\end{array}
\right. 
\label{even_odd}
\end{equation}
Hence, in addition to the \textit{overall} order of a term in the series,
we only need to keep track of the \textit{parity} of the number of times
each \textit{individual} bond is used. And, at the end the disorder average
for $J_b^n$ is simply $J^n$ for $n$ even and zero for $n$ odd.

We write $\mathcal{H}_1$ as
\begin{equation}
\mathcal{H}_1 = J \sum_{b=1}^B \widetilde{\mathcal{H}}_b
\end{equation}
where
\begin{equation}
\widetilde{\mathcal{H}}_b = \epsilon_b\, \sigma^x_{b_1} \, \sigma^x_{b_2} \,
\end{equation}
is the perturbation due to bond $b$, which has the effect of flipping the two
spins, $b_1$ and $b_2$, connected to it, i.e.
\begin{equation}
\text{flip}_b\, (S_{b_1}, S_{b_2}) =  (-S_{b_1}, -S_{b_2}) \, 
\label{flip}
\end{equation}
where the notation $\text{flip}_b$ means act with bond $b$ to flip the two
spins connected to it, and $S_{b_1}$ etc.~refers to the value of
$\sigma^x_{b_1}$ in the basis state being considered.

We note that the perturbation has no
diagonal matrix elements among the basis states and, in particular, 
$\langle 0|\widetilde{\mathcal{H}}_b |0\rangle = 0\, .$

We will develop perturbation theory for the ground state of the cluster in
powers of the $J_b$. The unperturbed ground state, $|0\rangle$, has all spins
along $z$ and has energy $E_0$. We denote a general unperturbed eigenstate by
$|\alpha\rangle$ and its energy by $E_\alpha$.
These will be our (normalized) basis states.

The perturbation expansion is in powers of $J$, but we also need to specify
\textit{which} bonds have been used. Acting with a bond flips both spins attached
to this bond, see Eq.~\eqref{flip}, so acting twice with the bond leaves the original state
unchanged. The result of acting
on the ground state with a product of bond terms can therefore be specified by
a set of bits, $l_i$, which
give the \textit{parity} of the number of times bond
$i$ has acted: $l_i = 0$ for an even number of times, and $l_i=1$ for an odd
number. Similarly, since the bonds take only the two values $\pm J$, see
Eq.~\eqref{Jb}, a
product of $n$ bonds, some of which might occur more than once, can also
be written in terms of the $l_i$ as 
\begin{equation}
\prod_k J_k =
J^n \left(\prod_{b=1}^B \epsilon_b^{l_b}\right) \, ,
\end{equation}
where we recall that $B$ is the number of bonds in the cluster.)
To make the notation more compact we write the
bits $\{l_b\}$ as a single integer $L \ (=0, 1, \cdots, 2^B-1)$ where
\begin{equation}
L = \sum_{b=1}^B 2^{b-1}\, l_b \, .
\label{L}
\end{equation}
In other words, $l_b$ is the $b$-th bit in the bitwise representation of $L$.
The unperturbed ground state $|0\rangle$ has $L=0$.

Acting with a set of bonds specified by an integer $L$ on $|0\rangle$ gives a 
basis state $|\alpha\rangle$ as follows:
\begin{equation}
|\alpha \rangle = \prod_{b} \left(\sigma^x_{b_1} \,
\sigma^x_{b_2}\right)^{l^L_b}
\, |0\rangle \, ,
\label{gen_alpha}
\end{equation}
where $l^L_b$ is the parity of the number of times (1, odd or 0, even) of
the number of times the
$b$-th bond appears in the graph represented by $L$.
Hence a given value of $L$ \textit{uniquely}
specifies a state.
However, the same $|\alpha\rangle$ can be obtained in
different ways by flipping different sets of bonds, so the mapping is \textit{many
to one}. 

As an example, for a set of bonds which form
closed loops each spin is flipped an even number of times so it maps to
the ground state, as does acting with no bonds, $L=0$.
We denote the set of bond strings which map to $|0\rangle$ by
$\{L_0\}$. A general bond string $L$ which maps to a basis state
$|\alpha\rangle$ is written as
\begin{equation}
L = L_0 \oplus \alpha 
\end{equation}
where $\oplus$ means bitwise addition modulo 2. For $L_0 = 0$ we have $\alpha
= L$ so, in this scheme, we label a spin state by
\textit{one representative string} of bonds $L$ which
maps to that state.
We emphasize that we are representing a basis state by one of the sets of
\textit{bonds} that maps to it from $|0\rangle$ (flipping both spins of
the bond), rather than by the \textit{spins themselves}.

One way to choose values of $\alpha$ and $L_0$ for a given bond string integer
$L$ is to run through consecutive integers starting from 0 up to
$2^B - 1$. For each integer
one determines the spin configuration. If it has not previously occurred then we
choose this $L$ to be the representative value for the spin state, i.e.~this
bond
state is specified by $\alpha = L$ and $L_0 = 0$. If the spin state has been
met before we associate $L$ with two integers, the value of $\alpha$ of that
(previously obtained) spin state, and $L_0$ where $L = L_0 \oplus \alpha$.
In this way, we determine necessary lookup tables $\alpha(L)$ and $L_0(L)$.
For an example see Appendix~\ref{sec:label_square}, especially Table
\ref{tab:states}.

We have already said that there $2^B$ values of $L$ which specify the
different bond strings. Since each bond flips two spins, only states in the
even subspace (i.e.~those with an even number of spins flipped)
will be  generated from the unperturbed ground state.
Hence the number of spin states generated will be half their total number,
i.e.~$2^{N-1}$. Consequently, the number of values of the bond strings $L_0$ which
map to the ground state (and which also gives the number of times \textit{each}
spin state is generated from all possible bond strings), is the ratio $2^B /
2^{N-1} = 2^{B+1-N}$. Note that this is $2^{C_n}$ where $C_n$ is the
cyclomatic number discussed in Sec.~\ref{sec:graphs}, i.e.~the number of
independent cycles in the graph.

We will also need information about the result of acting with an 
\textit{additional} bond $b$ on an
existing bond string. If the original bond string is
represented by an integer $L$ then adding one more bond
$b$ gives an integer $L'$ where 
\begin{equation}
L' = \text{flip}_b\, L,
\end{equation}
i.e.~$L'$ is trivially obtained from $L$ by flipping the $b$-th bit of $L$. Now $L = L_0 \oplus
\alpha$, and if $L' = L'_0 \oplus \alpha'$ we need to compute $\alpha'$ and
$L'_0$ from
\begin{equation}
(\alpha', L_0') = \text{flip}_b\,(\alpha, L_0) \, ,
\label{flipaLp}
\end{equation}
for each $(\alpha, L_0)$ and $b$. 
The precise mapping will depend on the
(arbitrary) choice of which of the possible $L$'s which map to state
$|\alpha\rangle$ is taken to be the ``representative'' value (i.e.~the one
which goes with $L_0 =0$). Given $L_0$ and $\alpha$ one first
determines $L$ from $L = L_0 \oplus \alpha$, then gets $L'$ from
$L' = \text{flip}_B\, L$, and finally uses the lookup tables, $\alpha' =
\alpha(L'), L'_0 = L_0(L')$ to get $\alpha'$ and $L'_0$.

\subsection{Series for the ground state energy and wavefunction}
\label{sec:gsewf}

We now set up the perturbation expansions for the ground state energy and ground state
wave function. 
The ground state wave function $|\psi_g\rangle$ will mix into the unperturbed
ground state $|0\rangle$ other basis states $|\alpha\rangle$ as follows, 
\begin{equation}
|\psi_g\rangle = |0\rangle + \sum_{n=1}^\infty J^n \sum_{L_0, \,\alpha \ne 0}
C_{n}^{\alpha\oplus L_0}\, G^{\alpha \oplus L_0}
|\alpha\rangle\, ,
\label{psi}
\end{equation}
where
\begin{equation}
G^{\alpha\oplus L_0} =  \left(\prod_{b=1}^B \epsilon_b^{l^{\alpha\oplus L_0}_b}\right) 
\label{GL}
\end{equation}
is a graph represented by an integer $L = \alpha\, \oplus\, L_0$
with $b$ bits, where 
$l^{\alpha\oplus L_0}_b = 1$ if the 
$b$-th interaction $\epsilon_b$
appears an odd number of times and 0 if it appears an even
number of times. Graphically,
an edge $b$ is present in the graph $G$ if the
interaction $\epsilon_b$ appears an odd number of times, otherwise it is absent.

To get the ground state energy, we start with the
unperturbed ground state
and act with perturbations which must lead, at the end, back to the
unperturbed ground
state.  Hence the only perturbations which contribute to the ground state
are those in the set $\{L_0\}$, i.e. 
\begin{equation}
E_g = E_0 + \sum_{n=1}^\infty J^n \sum_{L_0} e_{n}^{L_0}\,
G^{L_0} \, .
\label{E}
\end{equation}

We are free to choose the normalization of the ground state and do so by
requiring that the coefficient of the unperturbed ground state $|0\rangle$
in Eq.~\eqref{psi} is
precisely unity, so $|0\rangle$ is excluded from
the sum over states $|\alpha\rangle$ in Eq.~\eqref{psi}.
Expectation values therefore have to be
calculated from 
\begin{equation}
\langle \cdots \rangle = { \langle \psi_g | \cdots |\psi_g\rangle \over
\langle \psi_g | \psi_g\rangle}\, ,
\label{exp_value}
\end{equation}
where the normalizing denominator is given by
\begin{equation}
\langle \psi_g | \psi_g\rangle = 1 + \sum_{n=1}^\infty J^n \sum_{m=0}^n \sum_\alpha \sum_{L_0, L_0'}
C^{\alpha\oplus L_0}_m C^{\alpha\oplus L_0^{\prime}}_{n-m}  \,
G^{L_0^{\prime\prime}}
 \, ,
\end{equation}
in which $L_0^{\prime\prime} = L_0 \oplus L'$. This series can be written
\begin{equation}
\langle \psi_g | \psi_g\rangle = 1 + \sum_{n=1} J^n \sum_{L_0} U(n, L_0) G^{L_0} \, ,
\label{den}
\end{equation}
where we recall that the graph $G^{L_0}$ indicates which bonds have been used
to generate the term, and
\begin{equation}
U(n, L_0) = \sum_{m=0}^n \sum_\alpha \sum_{L_0'}
C^{\alpha\oplus L_0'}_m C^{\alpha\oplus L_0^{\prime}\oplus L_0}_{n-m} .
\end{equation}
It is necessary to keep track of which bonds have been used in
order to the final average over disorder. Note that in this expression we
only have to sum over bond configurations $L_0$ which map the ground state
back to the ground state.

The numerator in Eq.~\eqref{exp_value} can also be written in a form similar
to Eq.~\eqref{den} except that we are no longer restricted to bond
configurations which map the ground state to itself, so
\begin{equation}
\langle \psi_g | \cdots |\psi_g\rangle  = 
\sum_{n=0} J^n \sum_{L=0}^{2^B -1} V(n, L) G^{L} \, ,
\label{num}
\end{equation}
where the coefficients $V(n, L)$ can be determined from the ground state
expansion coefficients $C^{\alpha\oplus L_0}_n$ and a knowledge of the matrix
elements of the operator $(\cdots)$.
We discuss in
Appendix~\ref{sec:divide} how to divide the series in Eq.~\eqref{num} by that
in \eqref{den} efficiently.

The Schr\"odinger equation is
\begin{widetext}
\begin{multline}
\left\{\mathcal{H}_0 + J \sum_{b=1}^B \widetilde{\mathcal{H}}_b\right\}\,
\left\{|0\rangle + \sum_{n=1}^\infty J^n \sum_{L_0, \,\alpha \ne 0}
C_{n}^{\alpha \oplus L_0}\, 
G^{\alpha \oplus L_0}
|\alpha\rangle\right\} =
\\
\left\{E_0 + \sum_{n=1}^\infty J^n \sum_{L_0} e_{n}^{L_0} \,
G^{L_0}\right\}\,
\left\{|0\rangle + \sum_{n=1}^\infty J^n \sum_{L_0, \,\alpha \ne 0}
C_{n}^{\alpha\oplus L_0}\,
G^{\alpha \oplus L_0}
|\alpha\rangle\right\}\, .
\label{Sch}
\end{multline}
\end{widetext}

To proceed we first multiply both sides of Eq.~\eqref{Sch}
on the left by $\langle 0|$ and equate the
terms with the same order $n$ and bond graph $G^{\alpha\oplus L_0}$
on both sides. For $n = 0$ this trivially gives
$\langle 0| \mathcal{H}_0 |0\rangle = E_0 \langle 0|0 \rangle$.
For $n >0$ we have 
\begin{equation}
\sum_{b=1}^B
C_{n-1}^{\alpha\oplus L'_0}\, 
\langle 0| \sigma^x_{b_1} \, \sigma^x_{b_2} |\alpha\rangle
=  e_{n}^{L_0}
,
\label{CME}
\end{equation}
where
\begin{equation}
(\alpha, L'_0) = \text{flip}_b\, (0, L_0) \, .
\label{ab0}
\end{equation}
The matrix element is one so we have
\begin{equation}
e_{n}^{L_0}\ = \sum_{b=1}^B
C_{n-1}^{\alpha\oplus L'_0} .
\label{en}
\end{equation}

Next we multiply both sides of Eq.~\eqref{Sch} on the left by $\langle \gamma
|$ where $\gamma \ne 0$. This gives
\begin{multline}
(E_0 - E_\gamma) \, 
C_{n}^{\gamma\oplus L_0}\,
= 
\sum_{b=1}^B 
\langle \gamma| \sigma^x_{b_1} \, \sigma^x_{b_2} |\alpha\rangle
C_{n-1}^{\alpha\oplus L'_{0}}\,  \\
- \sum_{m=1}^{n-1}
\sum_{L'_0} e_{m}^{L'_0}\, 
C_{n-m}^{\gamma\oplus L_0 \oplus L'_0}\,
\, ,
\end{multline}
where, in the first term on the RHS,
\begin{equation}
(\alpha, L'_0) = \text{flip}_b\, (\gamma, L_0) . 
\label{abg}
\end{equation}
Simplifying we get
\begin{multline}
C_{n}^{\gamma\oplus L_0}
= {1 \over E_0 - E_\gamma}\, 
\\
\left\{
\sum_{b=1}^B C_{n-1}^{\alpha\oplus L'_{0}}
- \sum_{m=1}^{n-1}
\sum_{L''_0} e_{m}^{L''_0}\, 
C_{n-m}^{\gamma\oplus L_0\oplus L''_0}\,
\right\}
\, ,
\label{Cngamma}
\end{multline}
where we again note Eq.~\eqref{abg}.
In the second
term on the RHS of Eq.~\eqref{Cngamma}, the terms with $m=n$ and $m=0$ are 
not included, as can be seen by looking at the RHS of Eq.~\eqref{Sch}.
For the case of $n=1$, the second term on
the RHS of Eq.~\eqref{Cngamma} does not occur, and, in the first term, one
has $\alpha = L_0 = L'_0 = 0$ and 
$C^0_{0} = 1$.

To determine the terms in the expansion one proceeds as follows:
\begin{enumerate}
\item
Use Eq.~\eqref{Cngamma} with $n=1$ to determine the first order correction to
the wave function.

\item
\label{do_en}
Use Eq.~\eqref{en} with $n=2$ to get the second order contribution to the
energy.

\item
\label{do_psi}
Use Eq.~\eqref{Cngamma} with $n=2$ to determine the second order correction to
the wave function.

\item
Repeat steps \ref{do_en} and \ref{do_psi} to go to higher
orders.
\end{enumerate}

Note that in the RHS of
Eq.~\eqref{en}, and in the first term on the RHS of Eq.~\eqref{Cngamma}, for
each bond $b$ only
one value of $\alpha$ and $L'_0$ will contribute, given by Eqs.~\eqref{ab0} and
\eqref{abg} respectively.

In order to do the calculation we need to run through spin states (labeled
here by $\alpha$), and bond strings which map to the ground state  (labeled
here by $L_0$) \textit{consecutively}.
We therefore construct 
appropriate arrays to map from consecutive entries to $\alpha$ (and to $L_0$),
and for the inverse mapping from $\alpha$ (and $L_0$) to consecutive entries.

Having obtained the wavefunction coefficients in the expansion in
Eq.~\eqref{psi}, the series for each expectation value of interest is
obtained from Eq.~\eqref{exp_value}, in which the numerator and denominator
have the forms in Eq.~\eqref{num} and \eqref{den} respectively.
An efficient way to divide these series is given
in Appendix~\ref{sec:divide}.

In a spin glass the series for the expectation value has to be squared, see
Appendix~\ref{sec:divide}, and
then finally
averaged over the bond disorder. Since the expectation value of a bond is 0,
the only terms which survive after bond averaging are those where each bond
appears an even number of times, i.e.~those in which the bond string $L$, defined in
Eq.~\eqref{L}, is zero. 

\subsection{Series for the spin glass susceptibility}
\label{sec:chisg}

Next we want to compute the spin glass susceptibility. This can be done by
computing the change in energy to quadratic order in local,
magnetic fields which couple to $\sigma^x$.
Writing
\begin{equation}
E_g(\{h\}) = E_g(0) + {1\over 2}\sum_{i, j} \chi_{ij} h_i h_j + O(h^4) \, ,
\label{Eh}
\end{equation}
the spin glass susceptibility is given by
\begin{equation}
\chi_{SG} = \sum_{i, j} \left[\chi^2_{ij}\right]_\text{av}\, . 
\label{chisg}
\end{equation}
We will consider the field to act on just
two sites at a time, ``$i$" and ``$j"$.
Hence the Hamiltonian we consider is
\begin{equation}
\mathcal{H} = - \sum_k \sigma^z_k + J\, \sum_{b=1}^B \epsilon_b\,
\sigma^x_{b_1} \, \sigma^x_{b_2} + \left(h_i \sigma^x_i + h_j \sigma^x_j\right) \, .
\label{Hh}
\end{equation}
(Recall that $b$ is a bond and $\epsilon_b = \pm 1$.)

Up to now we have only needed to consider the even spin subspace, that is the
unperturbed ground state and all states obtained from it by flipping pairs of
spins. Recall from Sec.~\ref{sec:gsewf} that these can be characterized
by an integer $\alpha$ such that the $b$-th bit of $\alpha$ is 1 if bond $b$
is used to generate the spin state an odd number of times
and zero otherwise, see Eq.~\eqref{gen_alpha}. Several
integers $L$ (i.e.~several sets of bonds) can generate the same spin state
$\alpha$ and the graph $G$ of bonds is given by Eq.~\eqref{GL}.

How do we extend these ideas to the odd subspace?  First of all we write a
state in the even subspace as $|\alpha \rangle_e$ and one in the odd subspace
as $|\alpha \rangle_o$.  The unperturbed ground state in the even subspace,
$|0\rangle_e$, has all spins along the $+x$ direction. We define the
corresponding state in the odd subspace, $|0\rangle_o$, to be the state obtained
from $|0\rangle_e$ by flipping one of the spins coupled to one of the fields in
Eq.~\eqref{Hh}, let's say ``$i$", i.e.
\begin{equation}
\sigma^x_i \, |0 \rangle_e  = | 0 \rangle_o\, ,
\end{equation}
Other states in the odd subspace are then obtained in the same way as those in
the even subspace,
so
\begin{equation}
\sigma^x_i \, |\alpha \rangle_e  = | \alpha \rangle_o\, ,
\label{odd_states1}
\end{equation}
and similarly $\sigma^x_i \, |\alpha \rangle_o  = | \alpha \rangle_e$.

This representation is a convenient way to describe spin states obtained by
flipping the spin at site ``$i$" due to the field term in Eq.~\eqref{Hh} and
then acting with pair flips due to the bonds.  However,
we also will need to describe the spin state obtained by using the field to
flip the other site with a field in Eq.~\eqref{Hh}, namely ``$j$", and then act with a set
of pairwise bond flips. To do this we will need the state that has just sites
``$i$" and ``$j$" flipped. This will be represented by string of bonds between
these two sites,, which we will call $\alpha_{ij}$. Except for tree graphs there
will be several bond strings which do this, differing by closed-loop graphs
specified by $L_0$. We choose the one with $L_0 = 0$ (the empty graph).  

Hence, flipping spin ``$j$" and acting with bond graph $\alpha$ gives
\begin{multline}
\sigma^x_j \, |\alpha \rangle_e  =
\left(\prod_{b \in \alpha_{ij}} \sigma^x_{b_1} \, \sigma^x_{b_2}\right) 
\sigma^x_i \,|\alpha \rangle_e =\\ \sigma^x_i |\alpha \oplus \alpha_{ij} \rangle_e 
= |\alpha \oplus \alpha_{ij} \rangle_o \, . 
\label{odd_states2}
\end{multline}

We will compute the series expansion for the ground state energy and wave
function in powers of $J$ and up to second order in $h_i$ and $h_j$. The ground state
energy is an even function of the fields so the new piece is quadratic in 
$h_i$ and $h_j$. Each
term in the perturbation expansion for the GS energy
involves generating excited states from the unperturbed
ground state and ending up back in the unperturbed ground state.  Hence the
quadratic terms in the energy involve one of the following processes: (i) use $h_i$ to flip spin
``$i$" and flip it back, (ii) use $h_j$ to flip spin ``$j$" and flip it back,
(iii) use $h_i$ to flip spin ``$i$" and $h_j$ to flip spin ``$j$", and then
flip both these back with a bond string between them. The ground state wave
function will have terms linear in the fields as well as quadratic.

We therefore make the following ansatz.
\begin{widetext}
\begin{equation}
E_g = E_0 + \sum_{n=1}^\infty J^n \sum_{L_0} e_n^{L_0}\, G^{L_0} +
\sum_{n=0}^\infty J^n \sum_{L_0}
\left[
h_i^2
\, g^{(ii), L_0}_n\, G^{L_0} +
h_j^2
\, g^{(jj), L_0}_n\, G^{L_0} +
h_i h_j\,
g^{(ij), L_0}_n\, G^{L_0 \oplus \alpha_{ij}} \right] \, ,
\label{Eg_final}
\end{equation}
\begin{equation}
\begin{split}
|\psi_g\rangle = |0\rangle_e &+ \sum_{n=1}^\infty\, J^n\,
\sum_{\alpha \ne 0, L_0}
C_n^{\alpha\oplus L_0}\, G^{\alpha \oplus L_0} |\alpha\rangle_e \\
& + \sum_{n=0}^\infty\, J^n
\sum_{(\mathrm{all}\ \alpha), L_0} \left( h_i\,
A^{(i),\alpha\oplus L_0}_n\, G^{\alpha\oplus L_0} |\alpha\rangle_o + h_j\,
A^{(j),\alpha\oplus L_0}_n\, G^{\alpha\oplus L_0}
|\alpha\oplus\alpha_{ij}\rangle_o  
\right)
\\
& + \sum_{n=0}^\infty\, J^n
\sum_{\alpha\ne 0, L_0} \left( 
h_i^2\, D^{(ii),\alpha\oplus L_0}_n\, G^{\alpha\oplus L_0} +
h_j^2\, D^{(jj),\alpha\oplus L_0}_n\, G^{\alpha\oplus L_0} \right)
|\alpha\rangle_e  \\
& + \sum_{n=0}^\infty\, J^n
\sum_{\alpha\ne \alpha_{ij}, L_0} h_i\, h_j\,
D^{(ij),\alpha\oplus L_0}_n\, G^{\alpha \oplus L_0} 
|\alpha \oplus \alpha_{ij} \rangle_e \, .
\end{split}
\end{equation}
\end{widetext}

As for the zero field case in Sec.~\ref{sec:gsewf} we write down the
Schr\"odinger equation with these ansatzes, and project out the terms
separately by multiplying on the left by ${}_e\langle 0|$,
${}_e\langle \alpha|$ (for $\alpha\ne 0$), and ${}_o\langle \alpha|$.

After some algebra we find the following results.
For order $n=0$ (remember $n$ is the lower index)
all quantities are zero except for
\begin{equation}
C^0_0 = 1 \, , \ \ 
A^{(i),0}_0 = -{1\over 2} \, , \ \ 
g^{(ii), 0}_0 = -{1\over 2} \, , \ \  
D^{(ij),0}_0 = {1 \over 4}\, .
\end{equation}

The quantities which are independent of the $h_i$, namely $e_n^{L_0}$ and $C_n^{\alpha
\oplus L_0}$, have only to be calculated once, and the formulae for them were
already obtained in Eqs.~\eqref{en} and \eqref{Cngamma} in
Sec.~\ref{sec:gsewf} above.

The quantities involving one field $h_i$
have to be calculated $N$ times, once for each value of the site $i$ where
the field is applied. The description of the odd states is set up separately
for each value of $i$, such that $|0\rangle_o = \sigma^x_i |0\rangle_e$.
We find
\begin{equation}
g^{(ii), L_0}_n =
\sum_{b=1}^B D^{(ii),\alpha\oplus L_0'}_{n-1} +
A^{(i),L_0}_n\, ,
\label{gii_gen}
\end{equation}
\begin{multline}
A^{(i),\alpha\oplus L_0}_n = {1 \over E_0 - E^\mathrm{odd}_\alpha}
\\
\left\{
\sum_{b=1}^B A^{(i),\gamma\oplus L_0'}_{n-1} + C^{\alpha\oplus L_0}_n - \sum_{m=1}^n
\sum_{L_0'} e_m^{L_0'} A^{(i),\alpha \oplus L_0 \oplus L_0'}_{n-m}
\right\} \, ,
\label{A_gen}
\end{multline}


\begin{multline}
D^{(ii),\alpha\oplus L_0}_n = {1 \over E_0 - E_\alpha} \Biggl\{
\sum_{b=1}^B 
D^{(ii),\gamma\oplus L_0'}_{n-1} + A^{(i),\alpha\oplus L_0}_n
\\
-\sum_{m=0}^{n-1}\sum_{L_0'} g^{(ii), L_0'}_m\,
C^{\alpha\oplus L_0\oplus L_0'}_{n-m}
- \sum_{m=1}^{n}\sum_{L_0'} e_m^{L_0'}\,
D^{(ii),\alpha\oplus L_0\oplus L_0'}_{n-m} 
\Biggr\} .
\label{Dii_gen}
\end{multline}

Next we give expressions for quantities involving two fields $h_i$ and $h_j$. 
These have to be calculated $N(N-1)/2$ times,
once for each pair $i$ and $j$. The description of the odd states has to be
set up separately for each pair.  We find
\begin{equation}
g^{(ij), L_0}_n = \sum_{b=1}^B D^{(ij),
\alpha\oplus \alpha_{ij} \oplus L_0'}_{n-1} +
A^{(i),\alpha_{ij}\oplus L_0}_n +
A^{(j),\alpha_{ij}\oplus L_0}_n \, ,
\label{gij_gen}
\end{equation}
\begin{multline}
D^{(ij),\alpha\oplus L_0}_n = {1 \over E_0 - E_{\alpha\oplus\alpha_{ij}}}
\Biggl\{
\sum_{b=1}^B 
D^{(ij),\gamma\oplus L_0'}_{n-1}
+ A^{(i),\alpha\oplus L_0}_n \\
+ A^{(j),\alpha\oplus L_0}_n 
 -
\sum_{m=0}^{n-1}\sum_{L_0'} g^{(ij),L_0'}_m\,
C^{\alpha\oplus\alpha_{ij}\oplus L_0\oplus L_0'}_{n-m}
\\
-
\sum_{m=1}^{n}\sum_{L_0'} e^{L_0'}_m\,
D^{(ij),\alpha\oplus L_0\oplus L_0'}_{n-m} 
\Biggr\} \, .
\label{Dij_gen}
\end{multline}

The procedure to calculate the series iteratively is therefore as follows:
\begin{itemize}
\item
Compute the zero field quantities first. (Note that $C^0_0 = 1$.)
\begin{itemize}
\item
Use Eq.~\eqref{Cngamma} with $n=1$ to compute the first order correction to the wave
function.
\item
Use Eq.~\eqref{en}
followed by Eq.~\eqref{Cngamma} with $n= 2, 3, \cdots$ to compute
successive orders.
\end{itemize}
Equation \eqref{en} is not applied for $n=1$ because the RHS of that
equation vanishes in this case.
\item
Compute quantities involving one field. 
(Note that $A^{(i),0}_0 = -{1/ 2}, g^{(ii), 0}_0 = -{1/ 2}$.)\\
Use Eq.~\eqref{A_gen} followed by Eq.~\eqref{Dii_gen} followed by
Eq.~\eqref{gii_gen} iteratively
with $n = 1, 2, 3, \cdots$.
\item
Compute quantities involving two fields. (Note that $D_0^{(ij),0} = -1/4$.)\\
Use Eq.~\eqref{Dij_gen} followed by Eq.~\eqref{gij_gen} iteratively with $n=1, 2, 3,
\cdots$.
\end{itemize}


Our goal is to calculate the spin glass susceptibility $\chi_{SG}$ given
by Eq.~\eqref{chisg}.
Now the change in energy is related to the local susceptibilities $\chi_{ij}$
given by Eq.~\eqref{Eh}.
Comparing with Eq.~\eqref{Eg_final} we see that
\begin{align}
\chi_{ii} &= 2 \sum_{n=0}^\infty J^n \sum_{L_0}  g^{(ii), L_0}_n\, G^{L_0} , \\
\chi_{ij} &= \sum_{n=0}^\infty J^n \sum_{L_0}  g^{(ij), L_0}_n\,
G^{L_0\oplus\alpha_{ij}} , \quad (i \ne j), \\
\chi_{ji} &= \chi_{ij} \, .
\end{align}
We square these expressions, and then average over disorder which gives
except for the graph with no
bonds, $G^0$, because the
average of each bond, $J \epsilon_b$, is zero. Hence
\begin{align}
&[\chi_{ii}^2]_\text{av} =
4 \sum_{n=0}^\infty J^n \sum_{L_0}\sum_{m=0}^n g^{(ii), L_0}_m\, 
g^{(ii), L_0}_{n-m} \, ,
\\
&
[\chi_{ij}^2]_\text{av} = \sum_{n=0}^\infty J^n \sum_{L_0} 
\sum_{m=0}^n\, g^{(ij), L_0}_m\, g^{(ij), L_0}_{n-m} \, 
, \quad (i \ne j).
\end{align}
Summing these expression over sites gives $\chi_{SG}$ according to
Eq.~\eqref{chisg}.


\section{Finite Cluster Calculations for Ising spin-glasses in a classical field}
\label{sec:classical}

In this section, we consider weight calculations for
the Ising spin-glass in a classical field. The problem is defined
by the Hamiltonian:
\begin{equation}
\mathcal{H} = -\sum_{\langle i, j\rangle} J_{ij} S_i S_j - h\sum_{i=1}^N S_i
\, ,
\end{equation}
where the $S_i$ are Ising spins which take values $\pm 1$, and the
interactions $J_{ij}$ are \textit{quenched} random variables, again with a bimodal
distribution, $J_{ij} = \pm J$ with equal probability.

The key quantity of interest is the spin glass
susceptibility $\chiSG$ where
\begin{equation}
\chiSG = {1\over N} \sum_{i, j= 1}^N \left[ \bigl(\,\langle S_i S_j \rangle -
\langle S_i \rangle\, \langle S_j \rangle\,\bigr)^2 \right]_\mathrm{av} \, .
\end{equation}
For a fixed value of $h$ we expand the susceptibility in powers of
\begin{equation}
w = \tanh^2(J/T) \,.
\label{w_def}
\end{equation}
The coefficient of $w^n$ turns out to be a polynomial of order $2n +2$ in
\begin{equation}
u = \tanh^2(h/T),
\label{u}
\end{equation}
so
\begin{equation}
\chiSG(w, u) =\sum_{n=0}^\infty \,\left(\,  \sum_{m=0}^{2n+2} a_{n,m} u^m
\,\right)\, w^n \, .
\label{series}
\end{equation}
Our goal is to calculate the series coefficients for the spin-glass susceptibility on a
{\it finite} cluster with $B$ bonds and $N$ sites.

We begin with the partition function
\begin{align}
Z & \equiv \mathrm{Tr} \exp{(-\beta H)} \nonumber \\
& = \sum\limits_{\{S_i=\pm 1\}}
\exp{\Bigl(\beta \sum\limits_{\langle i,j\rangle } J_{ij} S_i S_j +
\beta h \sum\limits_i S_i \Bigr) }\, .
\end{align}
We use the relations
\begin{equation}
\exp{(\beta J_{ij} S_i S_j)}= \cosh{\beta J}\ ( 1 + v_{ij} S_i S_j),
\end{equation}
with $v_{ij}=\tanh{\beta J_{ij}}$, and
\begin{equation}
\exp{(\beta h S_i )}= \cosh{\beta h}\  ( 1 + b S_i ),
\end{equation}
with $b=\tanh{\beta h}$. 
Let us define
\begin{equation}
Z^\prime = {Z\over (\cosh{\beta J})^{B}(2\cosh{\beta h})^N},
\end{equation}
where, as stated above,
$B$ is the number of bonds and $N$ the number of sites in the cluster. Then,
\begin{equation}
Z^\prime
= { \mathrm{Tr} \prod_{\langle i,j\rangle } (1 + v_{ij} S_i S_j)
\prod_{i} (1+b S_i)\over 2^N}.
\end{equation}

The main task is to expand $Z^\prime$ in powers of the $v_{ij}$ and $b$. When the
two products are expanded the trace will be non-zero only if each spin occurs an even
number of times. In that case the trace will give a factor of $2^N$, canceling
that factor in the denominator. The first product over the bonds has $2^{B}$ terms
and each one combines uniquely with a suitable number of powers of $b$ from
the second term to give a non-zero contribution. Thus the trace results in
precisely $2^{B}$ terms. Thermal averages such as $\langle S_i S_j\rangle$ or
$\langle S_i\rangle$ can be expressed as a ratio of two such traces each
of which has $2^{B}$ terms.

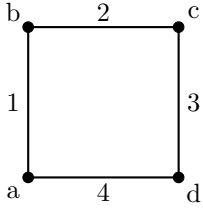
\begin{figure}
\begin{tikzpicture}

\filldraw[black] (-1,-1) circle (2pt);
\filldraw[black] (-1, 1) circle (2pt);
\filldraw[black] ( 1,-1) circle (2pt);
\filldraw[black] ( 1, 1) circle (2pt);

\node  (a) at (-1.2,-1.2) {a};
\draw [thick]  (-1,-1) --  (-1,  1);
\node  (1) at (-1.2, 0) {1};

\draw [thick]  (-1, 1) --  ( 1,  1);
\node  (b) at (-1.2, 1.2) {b};
\node  (2) at (0, 1.2) {2};

\draw [thick]  ( 1, 1) --  ( 1, -1);
\node  (c) at (1.2, 1.2) {c};
\node  (3) at (1.2, 0) {3};

\draw [thick]  ( 1, -1) -- (-1, -1);
\node  (d) at (1.2, -1.2) {d};
\node  (4) at (0, -1.2) {4};

\end{tikzpicture}
\caption{\label{fig:example}
Labeling of spins a, b, c and d (at the corners) and bonds (1, 2, 3, and 4) on
the edges, for the example in Appendix \ref{sec:label_square}
}
\end{figure}

These terms have a simple graphical representation. To illustrate that, we
consider an elementary square graph consisting of $4$ sites and $4$ bonds
shown in Fig.~\ref{fig:example}. The expressions for the numerator of $\langle S_a S_d\rangle$
and $\langle S_a\rangle$  and their common denominator are shown graphically
in Figs.~\ref{fig:zprime}--\ref{fig:s1}. A line on bond $j$ gives a factor of
$\tanh(\beta J_j)$, and we have defined 
\begin{equation}
b = \tanh(\beta h)\, .
\end{equation}
Note that in zero field there would have been only two
non-zero terms for the graphs in Figs.~\ref{fig:zprime} and \ref{fig:s1s2} (and non 
at all in Fig.~\ref{fig:s1}), but now $16$ terms contribute for all these graphs.
This is one of the main
sources of complexity in working with a non-zero field.

The explicit expressions for the graphs in
Figs.~\ref{fig:zprime}--\ref{fig:s1} are as follows. 
Labeling the bonds as in Fig.~\ref{fig:example}, and noting that $v_{i} \equiv
\tanh \beta J_i = \epsilon_i \tanh \beta J =\epsilon_i\, v$ where
\begin{equation}
v = \tanh (\beta J) 
\end{equation}
and $\epsilon_i = \pm 1$, the
expression for $Z^\prime$ in Fig.~\ref{fig:zprime} is 
\begin{multline}
Z^\prime = 1 +
v^4
\epsilon_1 \epsilon_2 \epsilon_3 \epsilon_4 
+ b^4  v^2
\left( \epsilon_2  \epsilon_4 + \epsilon_1 \epsilon_3 \right) 
+ \\
b^2 v \left( \epsilon_4 + \epsilon_2 + \epsilon_1 + \epsilon_3
\right) 
+ \\
b^2  v^2
\left(\epsilon_4 \epsilon_1 + \epsilon_1 \epsilon_2 + \epsilon_2 \epsilon_3 + \epsilon_3
\epsilon_4
\right) 
+ \\
b^2 v^3 \left(\epsilon_1 \epsilon_3 \epsilon_4 + \epsilon_1 \epsilon_2 \epsilon_3 +
\epsilon_1 \epsilon_2 \epsilon_4 + \epsilon_2 \epsilon_3 \epsilon_4
\right) .
\end{multline}

Similarly in Fig.~\ref{fig:s1s2}, the expression for $Z^\prime \langle S_a
S_d\rangle$ is
\begin{multline}
Z^\prime \langle S_a S_d\rangle = v \epsilon_4 +
v^3 \epsilon_1 \epsilon_2 \epsilon_3 
+ 
b^4
\left( v^3 
\epsilon_1 \epsilon_3 \epsilon_4 + v\,\epsilon_2
\right) + \\
b^2
\left(
1 +  v^4 \epsilon_1 \epsilon_2 \epsilon_3 \epsilon_4 
\right) + b^2 v
\left( \epsilon_1 + \epsilon_3
\right) + \\
b^2 v^2 \left(
\epsilon_4 \epsilon_1 + 
\epsilon_1 \epsilon_2 + 
\epsilon_2 \epsilon_3 + 
\epsilon_3 \epsilon_4 +
\epsilon_2 \epsilon_4 + 
\epsilon_1 \epsilon_3 
\right) 
+ \\
b^2  v^3 \left(
\epsilon_1 \epsilon_2 \epsilon_4 + 
\epsilon_2 \epsilon_3 \epsilon_4  
\right) \, .
\end{multline}
Finally the expression for $Z^\prime \langle S_a \rangle $ in Fig.~\ref{fig:s1} is
\begin{multline}
Z^\prime \langle S_a \rangle = b  + b\, v^4 \left(
1 + \epsilon_1 \epsilon_2 \epsilon_3 \epsilon_4 \right) + 
b\,  v\, \left(\epsilon_1 + \epsilon_4
\right) + \\
b\, v^2 \left( \epsilon_3 \epsilon_4 + \epsilon_1 \epsilon_2
\right) + 
b\, v^3 \left(\epsilon_1 \epsilon_2 \epsilon_3 + \epsilon_2 \epsilon_3 \epsilon_4
\right) + \\
b^3 v\, \left(\epsilon_2 + \epsilon_3\right) + 
b^3 v^2 \left(
\epsilon_4 \epsilon_1 + \epsilon_2 \epsilon_3 + \epsilon_2 \epsilon_4 +
\epsilon_1 \epsilon_3 
\right) + \\
b^3 v^3  \left(
\epsilon_4 \epsilon_1 \epsilon_2 + \epsilon_1 \epsilon_3 \epsilon_4
\right) \, .
\end{multline}

The products of the $\epsilon_i$ are visualized as a bond graph $G^L$, as
discussed in Sec.~\ref{sec:gsewf} and illustrated in Appendix
\ref{sec:label_square} for the square cluster in Fig.~\ref{fig:example}.
The series can be divided and multiplied as discussed in Appendix
\ref{sec:divide}, the only difference here being that we have a double series
in $v =\tanh(\beta J)$ and $b = \tanh(\beta h)$, but the methods of Appendix
\ref{sec:divide} are easily generalized to this case.

\begin{figure}
\begin{tikzpicture}
\node (a) at  (-4.5,0)  {\large +};
\filldraw[black] (-5.5,0.25) circle (1pt);
\filldraw[black] (-5.0,0.25) circle (1pt);
\filldraw[black] (-5.5,-0.25) circle (1pt);
\filldraw[black] (-5.0,-0.25) circle (1pt);
\draw [thick] (-4,-0.25) -- (-3.5,-0.25);
\draw [thick]  (-3.5, -0.25) -- (-3.5,.25);
\draw [thick]  (-3.5,.25) -- (-4,.25);
\draw [thick]  (-4,.25) -- (-4,-0.25);
\filldraw[black] (-3.5,0.25) circle (1pt);
\filldraw[black] (-4.0,0.25) circle (1pt);
\filldraw[black] (-3.5,-0.25) circle (1pt);
\filldraw[black] (-4.0,-0.25) circle (1pt);
\node (b) at  (-2.5,0)  {\large + $b^4$ \Large (};
\draw [thick] (-1.5,-0.25) -- (-1,-0.25);
\draw [thick]  (-1.5,.25) -- (-1,.25);
\filldraw[black] (-1.5,0.25) circle (1pt);
\filldraw[black] (-1.0,0.25) circle (1pt);
\filldraw[black] (-1.5,-0.25) circle (1pt);
\filldraw[black] (-1.0,-0.25) circle (1pt);
\node (c) at  (-0.5,0)  {\large +};
\draw [thick]  (0, -0.25) -- (0,.25);
\draw [thick]  (0.5,.25) -- (0.5,-0.25);
\filldraw[black] (0.5,0.25) circle (1pt);
\filldraw[black] (0.0,0.25) circle (1pt);
\filldraw[black] (0.5,-0.25) circle (1pt);
\filldraw[black] (0.0,-0.25) circle (1pt);
\node (c) at  (1.0,0)  {\Large )};

\node (c) at  (-4.75,-1)  {\large + $b^2$ \Large (};
\draw [thick] (-4.0,-1.25) -- (-3.5,-1.25);
\filldraw[black] (-3.5,-1.25) circle (1pt);
\filldraw[black] (-4.0,-1.25) circle (1pt);
\filldraw[black] (-3.5,-0.75) circle (1pt);
\filldraw[black] (-4.0,-0.75) circle (1pt);
\node (b) at  (-3.0,-1)  {\large + };
\draw [thick]  (-2.5,-.75) -- (-2.0,-.75);
\filldraw[black] (-2.0,-1.25) circle (1pt);
\filldraw[black] (-2.5,-1.25) circle (1pt);
\filldraw[black] (-2.0,-0.75) circle (1pt);
\filldraw[black] (-2.5,-0.75) circle (1pt);
\node (b) at  (-1.5,-1)  {\large + };
\draw [thick]  (-1.0, -1.25) -- (-1.0,-.75);
\filldraw[black] (-1.0,-1.25) circle (1pt);
\filldraw[black] (-0.5,-1.25) circle (1pt);
\filldraw[black] (-1.0,-0.75) circle (1pt);
\filldraw[black] (-0.5,-0.75) circle (1pt);
\node (b) at  (0,-1)  {\large + };
\draw [thick]  (1,-1.25) -- (1,-0.75);
\filldraw[black] (1.0,-1.25) circle (1pt);
\filldraw[black] (0.5,-1.25) circle (1pt);
\filldraw[black] (1.0,-0.75) circle (1pt);
\filldraw[black] (0.5,-0.75) circle (1pt);

\node (d) at  (-4.5,-2)  {\large + };
\draw [thick] (-4.0,-2.25) -- (-3.5,-2.25);
\draw [thick] (-4.0,-2.25) -- (-4.0,-1.75);
\filldraw[black] (-3.5,-2.25) circle (1pt);
\filldraw[black] (-4.0,-2.25) circle (1pt);
\filldraw[black] (-3.5,-1.75) circle (1pt);
\filldraw[black] (-4.0,-1.75) circle (1pt);
\node (b) at  (-3.0,-2)  {\large + };
\draw [thick]  (-2.5,-1.75) -- (-2.0,-1.75);
\draw [thick]  (-2.5,-1.75) -- (-2.5,-2.25);
\filldraw[black] (-2.0,-2.25) circle (1pt);
\filldraw[black] (-2.5,-2.25) circle (1pt);
\filldraw[black] (-2.0,-1.75) circle (1pt);
\filldraw[black] (-2.5,-1.75) circle (1pt);
\node (b) at  (-1.5,-2)  {\large + };
\draw [thick]  (-0.5, -2.25) -- (-0.5,-1.75);
\draw [thick]  (-1.0, -1.75) -- (-0.5,-1.75);
\filldraw[black] (-1.0,-2.25) circle (1pt);
\filldraw[black] (-0.5,-2.25) circle (1pt);
\filldraw[black] (-1.0,-1.75) circle (1pt);
\filldraw[black] (-0.5,-1.75) circle (1pt);
\node (b) at  (0,-2)  {\large + };
\draw [thick]  (1,-2.25) -- (1,-1.75);
\draw [thick]  (1,-2.25) -- (0.5,-2.25);
\filldraw[black] (1.0,-2.25) circle (1pt);
\filldraw[black] (0.5,-2.25) circle (1pt);
\filldraw[black] (1.0,-1.75) circle (1pt);
\filldraw[black] (0.5,-1.75) circle (1pt);

\node (d) at  (-4.5,-3)  {\large + };
\draw [thick] (-4.0,-3.25) -- (-3.5,-3.25);
\draw [thick] (-4.0,-3.25) -- (-4.0,-2.75);
\draw [thick] (-3.5,-3.25) -- (-3.5,-2.75);
\filldraw[black] (-3.5,-3.25) circle (1pt);
\filldraw[black] (-4.0,-3.25) circle (1pt);
\filldraw[black] (-3.5,-2.75) circle (1pt);
\filldraw[black] (-4.0,-2.75) circle (1pt);
\node (b) at  (-3.0,-3)  {\large + };
\draw [thick]  (-2.5,-2.75) -- (-2.0,-2.75);
\draw [thick]  (-2.5,-2.75) -- (-2.5,-3.25);
\draw [thick]  (-2.0,-2.75) -- (-2.0,-3.25);
\filldraw[black] (-2.0,-3.25) circle (1pt);
\filldraw[black] (-2.5,-3.25) circle (1pt);
\filldraw[black] (-2.0,-2.75) circle (1pt);
\filldraw[black] (-2.5,-2.75) circle (1pt);
\node (b) at  (-1.5,-3)  {\large + };
\draw [thick]  (-1.0, -3.25) -- (-1.0,-2.75);
\draw [thick]  (-1.0, -3.25) -- (-0.5,-3.25);
\draw [thick]  (-1.0, -2.75) -- (-0.5,-2.75);
\filldraw[black] (-1.0,-3.25) circle (1pt);
\filldraw[black] (-0.5,-3.25) circle (1pt);
\filldraw[black] (-1.0,-2.75) circle (1pt);
\filldraw[black] (-0.5,-2.75) circle (1pt);
\node (b) at  (0,-3)  {\large + };
\draw [thick]  (1,-3.25) -- (1,-2.75);
\draw [thick]  (0.5,-2.75) -- (1,-2.75);
\draw [thick]  (0.5,-3.25) -- (1,-3.25);
\filldraw[black] (1.0,-3.25) circle (1pt);
\filldraw[black] (0.5,-3.25) circle (1pt);
\filldraw[black] (1.0,-2.75) circle (1pt);
\filldraw[black] (0.5,-2.75) circle (1pt);
\node (e) at  (1.5,-3)  {\Large ) };

\end{tikzpicture}
\caption{\label{fig:zprime}
Expansion of $Z^\prime$ for the square graph in Fig.~\ref{fig:example}
in which $b= \tanh(\beta h)$. A line on bond $j$ gives a factor of
$\tanh(\beta J_j) = \epsilon_j \tanh(\beta J) = \epsilon_j v$.
Each point with an odd number of lines going to it has a
factor of $b$.
Note that only the first two terms contribute in zero field.
}
\end{figure}
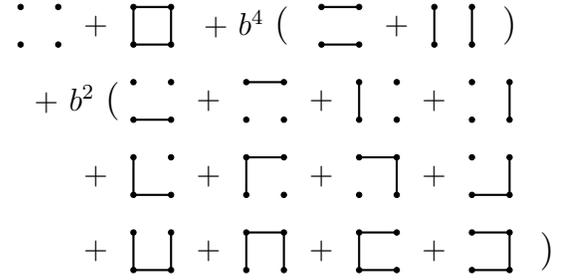

\begin{figure}
\begin{tikzpicture}

\draw [thick] (-5.5,-0.25) -- (-5.0,-0.25);
\filldraw[black] (-5.5,0.25) circle (1pt);
\filldraw[black] (-5.0,0.25) circle (1pt);
\filldraw[black] (-5.5,-0.25) circle (1.8pt);
\filldraw[black] (-5.0,-0.25) circle (1.8pt);

\node (c) at  (-4.5,0)  {\large +};

\draw [thick]  (-3.5, -0.25) -- (-3.5,.25);
\draw [thick]  (-3.5,.25) -- (-4,.25);
\draw [thick]  (-4,.25) -- (-4,-0.25);
\filldraw[black] (-3.5,0.25) circle (1pt);
\filldraw[black] (-4.0,0.25) circle (1pt);
\filldraw[black] (-3.5,-0.25) circle (1.8pt);
\filldraw[black] (-4.0,-0.25) circle (1.8pt);

\node (b) at  (-2.5,0)  {\large + $b^4$ \Large (};
\draw [thick] (-1.5,-0.25) -- (-1.5,.25);
\draw [thick] (-1,-0.25) -- (-1,.25);
\filldraw[black] (-1.5,0.25) circle (1pt);
\filldraw[black] (-1.0,0.25) circle (1pt);
\filldraw[black] (-1.5,-0.25) circle (1.8pt);
\filldraw[black] (-1.0,-0.25) circle (1.8pt);

\node (c) at  (-0.5,0)  {\large +};

\draw [thick]  (-1.5,-.25) -- (-1,-.25);
\draw [thick]  (0,.25) -- (0.5,0.25);
\filldraw[black] (0.5,0.25) circle (1pt);
\filldraw[black] (0.0,0.25) circle (1pt);
\filldraw[black] (0.5,-0.25) circle (1.8pt);
\filldraw[black] (0.0,-0.25) circle (1.8pt);

\node (c) at  (1.0,0)  {\Large )};

\node (c) at  (-4.75,-1)  {\large + $b^2$ \Large (};

\filldraw[black] (-3.5,-1.25) circle (1.8pt);
\filldraw[black] (-4.0,-1.25) circle (1.8pt);
\filldraw[black] (-3.5,-0.75) circle (1pt);
\filldraw[black] (-4.0,-0.75) circle (1pt);

\node (b) at  (-3.0,-1)  {\large + };
\draw [thick]  (-2.5,-.75) -- (-2.0,-.75);
\draw [thick]  (-2.0,-.75) -- (-2.0,-1.25);
\draw [thick]  (-2.5,-1.25) -- (-2.0,-1.25);
\draw [thick]  (-2.5,-1.25) -- (-2.5,-.75);
\filldraw[black] (-2.0,-1.25) circle (1.8pt);
\filldraw[black] (-2.5,-1.25) circle (1.8pt);
\filldraw[black] (-2.0,-0.75) circle (1pt);
\filldraw[black] (-2.5,-0.75) circle (1pt);

\node (b) at  (-1.5,-1)  {\large + };
\draw [thick]  (-1.0, -1.25) -- (-1.0,-.75);
\filldraw[black] (-1.0,-1.25) circle (1.8pt);
\filldraw[black] (-0.5,-1.25) circle (1.8pt);
\filldraw[black] (-1.0,-0.75) circle (1pt);
\filldraw[black] (-0.5,-0.75) circle (1pt);

\node (b) at  (0,-1)  {\large + };
\draw [thick]  (1,-1.25) -- (1,-0.75);
\filldraw[black] (1.0,-1.25) circle (1.8pt);
\filldraw[black] (0.5,-1.25) circle (1.8pt);
\filldraw[black] (1.0,-0.75) circle (1pt);
\filldraw[black] (0.5,-0.75) circle (1pt);

\node (d) at  (-4.5,-2)  {\large + };
\draw [thick] (-4.0,-2.25) -- (-3.5,-2.25);
\draw [thick] (-4.0,-2.25) -- (-4.0,-1.75);
\filldraw[black] (-3.5,-2.25) circle (1.8pt);
\filldraw[black] (-4.0,-2.25) circle (1.8pt);
\filldraw[black] (-3.5,-1.75) circle (1pt);
\filldraw[black] (-4.0,-1.75) circle (1pt);

\node (b) at  (-3.0,-2)  {\large + };
\draw [thick]  (-2.5,-1.75) -- (-2.0,-1.75);
\draw [thick]  (-2.5,-1.75) -- (-2.5,-2.25);
\filldraw[black] (-2.0,-2.25) circle (1.8pt);
\filldraw[black] (-2.5,-2.25) circle (1.8pt);
\filldraw[black] (-2.0,-1.75) circle (1pt);
\filldraw[black] (-2.5,-1.75) circle (1pt);

\node (b) at  (-1.5,-2)  {\large + };
\draw [thick]  (-0.5, -2.25) -- (-0.5,-1.75);
\draw [thick]  (-1.0, -1.75) -- (-0.5,-1.75);
\filldraw[black] (-1.0,-2.25) circle (1.8pt);
\filldraw[black] (-0.5,-2.25) circle (1.8pt);
\filldraw[black] (-1.0,-1.75) circle (1pt);
\filldraw[black] (-0.5,-1.75) circle (1pt);

\node (b) at  (0,-2)  {\large + };
\draw [thick]  (1,-2.25) -- (1,-1.75);
\draw [thick]  (1,-2.25) -- (0.5,-2.25);
\filldraw[black] (1.0,-2.25) circle (1.8pt);
\filldraw[black] (0.5,-2.25) circle (1.8pt);
\filldraw[black] (1.0,-1.75) circle (1pt);
\filldraw[black] (0.5,-1.75) circle (1pt);

\node (d) at  (-4.5,-3)  {\large + };
\draw [thick] (-4.0,-3.25) -- (-3.5,-3.25);
\draw [thick] (-4.0,-2.75) -- (-3.5,-2.75);
\filldraw[black] (-3.5,-3.25) circle (1.8pt);
\filldraw[black] (-4.0,-3.25) circle (1.8pt);
\filldraw[black] (-3.5,-2.75) circle (1pt);
\filldraw[black] (-4.0,-2.75) circle (1pt);

\node (b) at  (-3.0,-3)  {\large + };
\draw [thick]  (-2.5,-2.75) -- (-2.5,-3.25);
\draw [thick]  (-2.0,-2.75) -- (-2.0,-3.25);
\filldraw[black] (-2.0,-3.25) circle (1.8pt);
\filldraw[black] (-2.5,-3.25) circle (1.8pt);
\filldraw[black] (-2.0,-2.75) circle (1pt);
\filldraw[black] (-2.5,-2.75) circle (1pt);

\node (b) at  (-1.5,-3)  {\large + };
\draw [thick]  (-1.0, -3.25) -- (-1.0,-2.75);
\draw [thick]  (-1.0, -3.25) -- (-0.5,-3.25);
\draw [thick]  (-1.0, -2.75) -- (-0.5,-2.75);
\filldraw[black] (-0.5,-3.25) circle (1.8pt);
\filldraw[black] (-1.0,-3.25) circle (1.8pt);
\filldraw[black] (-0.5,-2.75) circle (1pt);
\filldraw[black] (-1.0,-2.75) circle (1pt);

\node (b) at  (0,-3)  {\large + };
\draw [thick]  (1,-3.25) -- (1,-2.75);
\draw [thick]  (0.5,-2.75) -- (1,-2.75);
\draw [thick]  (0.5,-3.25) -- (1,-3.25);
\filldraw[black] (1.0,-3.25) circle (1.8pt);
\filldraw[black] (0.5,-3.25) circle (1.8pt);
\filldraw[black] (1.0,-2.75) circle (1pt);
\filldraw[black] (0.5,-2.75) circle (1pt);

\node (e) at  (1.5,-3)  {\Large ) };

\end{tikzpicture}
\caption{\label{fig:s1s2}
Expansion of $Z^\prime \langle S_a S_d \rangle $ for the square graph in
Fig.~\ref{fig:example}(a) in which $b= \tanh(\beta h)$. The two spins $S_a$
and $S_d$ are on the lower horizontal bond of the square and are indicated
by the larger circles.
There is a factor of $b$ for each small circle
which has an odd number of lines going to it, and also a factor of $b$
for each large circle with
an even number of lines going to it.
Note again that only
the first two terms contribute in zero field.
}
\end{figure}
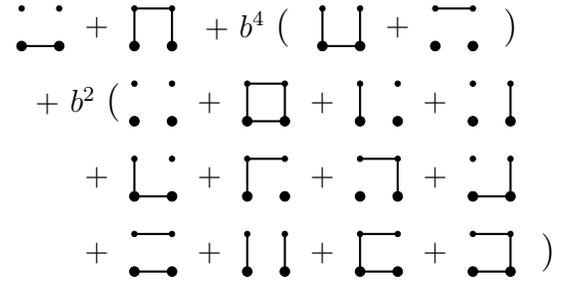

\begin{figure}
\begin{tikzpicture}

\node (c) at  (-5.0,0)  {\large \ $b$ \ \Large (};
\filldraw[black] (-3.5,-0.25) circle (1pt);
\filldraw[black] (-3.5, 0.25) circle (1pt);
\filldraw[black] (-4.0,-0.25) circle (1.8pt);
\filldraw[black] (-4.0, 0.25) circle (1pt);
\node (cc) at (-3.0, 0) {\large +};

\draw [thick]  (-2.0, -0.25) -- (-2.0,.25);
\draw [thick]  (-2.0,.25) -- (-2.5,.25);
\draw [thick]  (-2.5,.25) -- (-2.5,-0.25);
\draw [thick]  (-2.0,-.25) -- (-2.5,-0.25);
\filldraw[black] (-2.5,-0.25) circle (1.8pt);
\filldraw[black] (-2.5, 0.25) circle (1pt);
\filldraw[black] (-2.0,-0.25) circle (1pt);
\filldraw[black] (-2.0, 0.25) circle (1pt);

\node (b) at  (-1.5,0)  {\large + };
\draw [thick] (-1.0,-0.25) -- (-0.5,-.25);
\filldraw[black] (-0.5,-0.25) circle (1pt);
\filldraw[black] (-0.5, 0.25) circle (1pt);
\filldraw[black] (-1.0,-0.25) circle (1.8pt);
\filldraw[black] (-1.0, 0.25) circle (1pt);

\node (c) at  (0,0)  {\large +};
\draw [thick]  (0.5,-.25) -- (0.5,0.25);
\filldraw[black] (0.5,-0.25) circle (1.8pt);
\filldraw[black] (0.5, 0.25) circle (1pt);
\filldraw[black] (1.0,-0.25) circle (1pt);
\filldraw[black] (1.0, 0.25) circle (1pt);

\node (c) at  (-4.75,-1)  {\large + };
\draw [thick]  (-3.5,-1.25) -- (-4.0,-1.25);
\draw [thick]  (-3.5,-.75) -- (-3.5,-1.25);
\filldraw[black] (-3.5,-1.25) circle (1pt);
\filldraw[black] (-3.5,-0.75) circle (1pt);
\filldraw[black] (-4.0,-0.75) circle (1pt);
\filldraw[black] (-4.0,-1.25) circle (1.8pt);
 
\node (b) at  (-3.0,-1)  {\large + };
\draw [thick]  (-2.5,-.75) -- (-2.0,-.75);
\draw [thick]  (-2.5,-1.25) -- (-2.5,-.75);
\filldraw[black] (-2.5,-1.25) circle (1.8pt);
\filldraw[black] (-2.5, -0.75) circle (1pt);
\filldraw[black] (-2.0,-1.25) circle (1pt);
\filldraw[black] (-2.0, -0.75) circle (1pt);

\node (b) at  (-1.5,-1)  {\large + };
\draw [thick]  (-1.0, -1.25) -- (-1.0,-.75);
\draw [thick]  (-0.5, -.75) -- (-1.0,-.75);
\draw [thick]  (-0.5, -1.25) -- (-0.5,-.75);
\filldraw[black] (-0.5,-0.75) circle (1pt);
\filldraw[black] (-0.5,-1.25) circle (1pt);
\filldraw[black] (-1.0,-0.75) circle (1pt);
\filldraw[black] (-1.0,-1.25) circle (1.8pt);

\node (b) at  (0,-1)  {\large + };
\draw [thick]  (1,-1.25) -- (1,-0.75);
\draw [thick]  (0.5,-1.25) -- (1.0,-1.25);
\draw [thick]  (0.5,-.75) -- (1,-0.75);
\filldraw[black] (0.5,-1.25) circle (1.8pt);
\filldraw[black] (0.5,-0.75) circle (1pt);
\filldraw[black] (1.0,-1.25) circle (1pt);
\filldraw[black] (1.0,-0.75) circle (1pt);

\node (b) at  (1.5,-1)  {\Large ) };

\node (d) at  (-5.0,-2)  {\large $+\ b^3 \ $ \Large ( };
\draw [thick] (-4.0,-1.75) -- (-3.5,-1.75);
\filldraw[black] (-3.5,-2.25) circle (1pt);
\filldraw[black] (-3.5,-1.75) circle (1pt);
\filldraw[black] (-4.0,-1.75) circle (1pt);
\filldraw[black] (-4.0,-2.25) circle (1.8pt);

\node (b) at  (-3.0,-2)  {\large + };
\draw [thick]  (-2.0,-1.75) -- (-2.0,-2.25);
\filldraw[black] (-2.5,-2.25) circle (1.8pt);
\filldraw[black] (-2.5, -1.75) circle (1pt);
\filldraw[black] (-2.0,-2.25) circle (1pt);
\filldraw[black] (-2.0, -1.75) circle (1pt);

\node (b) at  (-1.5,-2)  {\large + };
\draw [thick]  (-1.0, -2.25) -- (-1.0,-1.75);
\draw [thick]  (-1.0, -2.25) -- (-0.5,-2.25);
\filldraw[black] (-0.5,-1.75) circle (1pt);
\filldraw[black] (-0.5,-2.25) circle (1pt);
\filldraw[black] (-1.0,-1.75) circle (1pt);
\filldraw[black] (-1.0,-2.25) circle (1.8pt);

\node (b) at  (0,-2)  {\large + };
\draw [thick]  (1,-2.25) -- (1,-1.75);
\draw [thick]  (1,-1.75) -- (0.5,-1.75);
\filldraw[black] (0.5,-2.25) circle (1.8pt);
\filldraw[black] (0.5,-1.75) circle (1pt);
\filldraw[black] (1.0,-2.25) circle (1pt);
\filldraw[black] (1.0,-1.75) circle (1pt);

\node (d) at  (-4.5,-3)  {\large + };
\draw [thick] (-4.0,-3.25) -- (-3.5,-3.25);
\draw [thick] (-4.0,-2.75) -- (-3.5,-2.75);
\filldraw[black] (-3.5,-3.25) circle (1pt);
\filldraw[black] (-3.5,-2.75) circle (1pt);
\filldraw[black] (-4.0,-2.75) circle (1pt);
\filldraw[black] (-4.0,-3.25) circle (1.8pt);

\node (b) at  (-3.0,-3)  {\large + };
\draw [thick]  (-2.5,-2.75) -- (-2.5,-3.25);
\draw [thick]  (-2.0,-2.75) -- (-2.0,-3.25);
\filldraw[black] (-2.5,-3.25) circle (1.8pt);
\filldraw[black] (-2.5, -2.75) circle (1pt);
\filldraw[black] (-2.0,-3.25) circle (1pt);
\filldraw[black] (-2.0, -2.75) circle (1pt);

\node (b) at  (-1.5,-3)  {\large + };
\draw [thick]  (-1.0, -3.25) -- (-1.0,-2.75);
\draw [thick]  (-1.0, -3.25) -- (-0.5,-3.25);
\draw [thick]  (-1.0, -2.75) -- (-0.5,-2.75);
\filldraw[black] (-0.5,-2.75) circle (1pt);
\filldraw[black] (-0.5,-3.25) circle (1pt);
\filldraw[black] (-1.0,-2.75) circle (1pt);
\filldraw[black] (-1.0,-3.25) circle (1.8pt);

\node (b) at  (0,-3)  {\large + };
\draw [thick]  (1,-3.25) -- (1,-2.75);
\draw [thick]  (0.5,-2.75) -- (0.5,-3.25);
\draw [thick]  (0.5,-3.25) -- (1,-3.25);
\filldraw[black] (0.5,-3.25) circle (1.8pt);
\filldraw[black] (0.5,-2.75) circle (1pt);
\filldraw[black] (1.0,-3.25) circle (1pt);
\filldraw[black] (1.0,-2.75) circle (1pt);

\node (e) at  (1.5,-3)  {\Large ) };

\end{tikzpicture}
\caption{\label{fig:s1}
Expansion of $Z^\prime \langle S_a\rangle $
for the square graph in Fig.~\ref{fig:example}
in which $b= \tanh(\beta h)$.
The spin $S_a$ is the lower left spin,
indicated by the larger circle.
There is a factor of $b$ for each small circle
which has an odd number of lines going to it, and also
a factor of $b$ for each large circle with
an even number of lines going to it.
Note that $Z^\prime \langle S_a\rangle $
is odd in the field and so vanishes identically in zero field.
}
\end{figure}

\section{Conclusions} 
In this manuscript, we have discussed a linked cluster based method for
calculating series expansions for Ising spin-glasses in a classical
(longitudinal) field and a
quantum (transverse) field. These expansions require \textit{all}
connected clusters (which we also denote by graphs)
that can be embedded on the lattice. The calculations require the counting and
enumeration of such graphs on the lattice, along with the weight calculations
for the quantities being computed, which are expanded
as series expansions for each graph.

We have discussed a method that takes a small list of graphs with no free ends
and obtains a complete list of all graphs by using the computer to
automatically generate suitable relations between lattice constants of graphs
with a free-end added to an existing graph thus avoiding the need for an
explicit enumeration of all lattice embeddings. Using available counts for no
free-end graphs in $d=2$ and $d=3$ we have generated all graphs in these cases
to $14$-th order. In higher than $3d$, we have only generated these counts to
$10$-th order using previously available star-graph counts. In future, it
should be possible to extend these counts in higher dimensions to at least
$14$-th order using the results of Brooks-Harris, Aharony 
and collaborators \cite{klein:91,daboul:04}.

We have discussed weight calculations for classical and quantum cases. The
$\pm J$ model provides great simplification in these calculations 
since
we only need to keep track of the
odd-even-ness of each bond before the final disorder averaging,
see Eq.~\eqref{even_odd}. This allows
for a simple graphical representation of the expansion that can be efficiently
dealt with on the computer. We find that the weight calculations for the
transverse-field case at $T=0$ are particularly efficient so that the
calculations are primarily limited by our ability to generate counts for no
free-end graphs. Thus, it should be possible to extend the series especially
in dimensionality greater than or equal to four when such graph counts are
available. By contrast, the finite temperature weight calculations for
classical spin-glasses in a field are much more time consuming, and it would be
difficult to extend these series to much higher orders using the present
methods.


\begin{acknowledgments}
One of us (APY) would like to thank the hospitality of the Indian Institute of
Science, Bangalore and the support of a DST-IISc Centenary Chair
Professorship.  He is particularly grateful for stimulating discussions with
H.~Krishnamurthy which initiated this project. The work of RRPS is supported
in part by US NSF grant number DMR-1306048. 
\end{acknowledgments}

\appendix
\section{An example of labeling states}
\label{sec:label_square}
We
illustrate the method of labeling states by considering the example of a
single square shown in Fig.~\ref{fig:example}. We label the bonds by 1, 2, 3
and 4, and the spins by a, b, c and d. The ground state has all spins up,
$|\!\uparrow \uparrow \uparrow \uparrow \rangle$.

The sequence in the spin labeling is a, b, c and d from right to left, so,
for example, the state
with spin c flipped is $|\!\uparrow \downarrow \uparrow \uparrow \rangle$.
Bond 1 is between spins a and b, bond 2 between spins b and c etc. We
represent the graph of bonds which are used to generate the spin states by an
integer $L$, where each bit of $L$ corresponds to a bond, and is 1 if that
bond is used to flip the two spins attached to it and 0 otherwise. 
The bits of $L$ are bonds 4, 3, 2 and 1 from left to right. Hence acting with
bond 3, which has
bit representation $L=0100$,
i.e.~$L=4$,
flips spins c and d and so gives state $|\!\downarrow
\downarrow \uparrow \uparrow \rangle$.

In this example, which has a single loop, each spin state can be generated by
two bond graphs. For example, the
ground state is represented by
the two bond integers $L=0$ and $15$ with bit representations
$L=0000$ and $L=1111$.
The lookup tables $\alpha(L)$ and $L_0(L)$, needed in the
computations, are shown in columns four and five of Table \ref{tab:states},
and are obtained as described in the figure caption. The spin configurations
for each $\alpha$, also needed in the computations, are shown in column three.

\begin{table}
\caption{\label{tab:states}
Labeling of states for the graph consisting of a single square shown in
Fig.~\ref{fig:example}. Note that in this graph, which has a single cycle,
each spin state is generated in two ways by acting with bonds.  Increasing $L$
from 0, we label the state by $\alpha = L$ and $L_0 = 0$ as long as the state
has not been found before. If the state has been found before we associate this
bond integer $L$
with the same value of $\alpha$ as found before but now $L_0 = 15$ (which
corresponds to the single closed loop of bonds for this graph).  In all cases
one has
$L = \alpha \oplus L_0$ where $\oplus$ is bitwise addition modulo 2.
Columns four and five of this table give us, for this graph, the lookup tables $\alpha(L)$
and $L_0(L)$ which are needed in the computations.  In this
way we can relate a set of bonds, specified by an integer $L$ to the 
label $\alpha$ for the spin state obtained by acting with those bonds on the
unperturbed ground state. We also need to store the spin configuration for each
value of $\alpha$ (column three).
}
\begin{tabular*}{\columnwidth}{@{\extracolsep{\fill}} |r|c|c|r|r|}
\hline
$L$ & bits of $L$ & spin state & $\alpha$ & $L_0$ \\
\hline\hline
$0$  & $0000 $ &$|\!\uparrow\uparrow\uparrow\uparrow\rangle$ & $0$ & $0$ \\
$1$  & $0001 $ &$|\!\uparrow\uparrow\downarrow\downarrow\rangle$ & $1$ & $0$ \\
$2$  & $0010 $ &$|\!\uparrow\downarrow\downarrow\uparrow\rangle$ & $2$ & $0$ \\
$3$  & $0011 $ &$|\!\uparrow\downarrow\uparrow\downarrow\rangle$ & $3$ & $0$ \\
$4$  & $0100 $ &$|\!\downarrow\downarrow\uparrow\uparrow\rangle$ & $4$ & $0$ \\
$5$  & $0101 $ &$|\!\downarrow\downarrow\downarrow\downarrow\rangle$ & $5$ & $0$ \\
$6$  & $0110 $ &$|\!\downarrow\uparrow\downarrow\uparrow\rangle$ & $6$ & $0$ \\
$7$  & $0111 $ &$|\!\downarrow\uparrow\downarrow\uparrow\rangle$ & $7$ & $0$ \\
$8$  & $1000 $ &$|\!\downarrow\uparrow\uparrow\downarrow\rangle$ & $7$ & $15$ \\
$9$  & $1001 $ &$|\!\downarrow\uparrow\downarrow\uparrow\rangle$ & $6$ & $15$ \\
$10$ & $1010 $ &$|\!\downarrow\downarrow\downarrow\downarrow\rangle$ & $5$ & $15$ \\
$11$ & $1011 $ &$|\!\downarrow\downarrow\uparrow\uparrow\rangle$ & $4$ & $15$ \\
$12$ & $1100 $ &$|\!\uparrow\downarrow\uparrow\downarrow\rangle$ & $3$ & $15$ \\
$13$ & $1101 $ &$|\!\uparrow\downarrow\downarrow\uparrow\rangle$ & $2$ & $15$ \\
$14$ & $1110 $ &$|\!\uparrow\uparrow\downarrow\downarrow\rangle$ & $1$ & $15$ \\
$15$ & $1111 $ &$|\!\uparrow\uparrow\uparrow\uparrow\rangle$ & $0$ & $15$ \\
\hline
\end{tabular*}
\end{table}

\section{Dividing Series}
\label{sec:divide}

Suppose we have two series of a single variable $x$, $f(x) = 
\sum_{n=0}^\infty a_n x^n$ and $g(x) = \sum_{n=0}^\infty b_n x^n$ and we want
the ratio,
$h(x) = \sum_{n=0}^\infty c_n x^n$, i.e.
\begin{equation}
\sum_{n=0}^\infty c_n x^n = {\sum_{n=0}^\infty a_n x^n \over \sum_{n=0}^\infty
b_n x^n} \, .
\label{series_ratio}
\end{equation}
We are given the $a_n$ and $b_n$ and want the $c_n$.
We assume without loss of generality that $b_0 = 1$. The simplest way to
compute the ratio in Eq.~\eqref{series_ratio} is to multiply both sides by the
denominator of the RHS, i.e.
\begin{equation}
(\sum_{n=0}^\infty b_n x^n ) \,
(\sum_{n=0}^\infty c_n x^n ) = 
\sum_{n=0}^\infty a_n x^n \, .
\end{equation}
Equating coefficients of powers of $x$ on both sides
one obtains the recursive equation
\begin{equation}
c_n = a_n - \sum_{k=1}^{n} b_k\, c_{n-k}\, \quad (n >0) ,
\label{cn}
\end{equation}
with $c_0 = a_0$. Using Eq.~\eqref{cn} for $n=1, 2, 3, \cdots$ in order,
determines the coefficients $c_n$.

In the present case, the series have an extra parameter, the integer $L$
representing the bonds that were used to generate this term, see
Eqs.~\eqref{exp_value},\eqref{num} and \eqref{den}. In other words we
have to generalize the above to
\begin{align}
f(x) &= \sum_{n=0}^\infty \sum_{L=0}^{2^B-1} a_{n,L} x^n, \nonumber\\
g(x) &= \sum_{n=0}^\infty \sum_{L=0}^{2^B-1} b_{n,L} x^n, \nonumber \\
h(x) \equiv f(x)/g(x) &= \sum_{n=0}^\infty \sum_{L=0}^{2^B-1} c_{n,L} x^n ,
\end{align}
where $b_{0,0} = 1$ and $b_{0,L} = L$ for $L \ne 0$.
Proceeding as
before, and noting that the factors of $L$ are combined using bitwise addition
modulo 2, we have
\begin{equation}
c_{n, L} = a_{n, L} - \sum_{k=1}^n \sum_{L'=0}^{2^B-1} b_{k, L'}\, c_{n-k, L' \oplus L}\, .
\end{equation}
We also have to multiply
series 
which is easily done by the above methods. The final stage is to
square the correlation functions and average over disorder, which means that
only the resulting term with $L=0$ survives.

\bibliography{refs,comments}

\begin{thebibliography}{25}
\bibitem[{int()}]{intensive}
\bibinfo{note}{Note that the since the lattice constant is the number of
  embeddings \textit{per site}, our calculation gives the intensive quantity
  $P/N$ even though $P(c)$ is the \textit{total} value of $P$ for cluster $c$.}
\end{thebibliography}

\end{document}